\documentclass{aastex}
\newfont{\rmsmall}{cmr10 scaled 900}
\newcommand{\ta}{T_{\mathrm{A}}}

\newcommand{\tat}{T_{\mathrm{at}}}

\newcommand{\tsky}{T_{\mathrm{sky}}}
\newcommand{\tsurf}{T_{\mathrm{surf}}}
\newcommand{\Tb}{T_{\mathrm{B}}}
\newcommand{\etal}{{\it et al.}}
\newcommand{\ci}{C~{\rmsmall I}}
\newcommand{\gs}{{\mathrel{\raise0.35ex\hbox{$\scriptstyle >$}\kern-0.6em 
\lower0.40ex\hbox}{{$\scriptstyle \sim$}}}}
\newcommand{\ls}{{\mathrel{\raise0.35ex\hbox{$\scriptstyle <$}\kern-0.6em 
\lower0.40ex\hbox}{{$\scriptstyle \sim$}}}}

\slugcomment{Submitted to the {\it PASP}, 5 October 2000}
\shortauthors{Stark et al.}
\shorttitle{The AST/RO Instrument}

\begin{document}

\title{THE ANTARCTIC SUBMILLIMETER TELESCOPE 
AND REMOTE OBSERVATORY (AST/RO)}

\author{Antony A. Stark\altaffilmark{1,7}, 
John Bally\altaffilmark{2,7}, 
Simon P. Balm\altaffilmark{1}, 
T. M. Bania\altaffilmark{3}, 
Alberto~D.~Bolatto\altaffilmark{3}, 
Richard~A.~Chamberlin\altaffilmark{3,4}, 
Gregory~Engargiola\altaffilmark{5,6}, 
Maohai~Huang\altaffilmark{3}, 
James~G.~Ingalls\altaffilmark{3,9},
Karl~Jacobs\altaffilmark{8}, 
James~M.~Jackson\altaffilmark{3}, 
Jacob~W.~Kooi\altaffilmark{9},
Adair~P.~Lane\altaffilmark{1}, 
K.-Y.~Lo\altaffilmark{5}, 
Rodney~D.~Marks\altaffilmark{1},
Christopher~L.~Martin\altaffilmark{1},
Dennis~Mumma\altaffilmark{7}, 
Roopesh~Ojha\altaffilmark{1},
Rudolf~Schieder\altaffilmark{8},
Johannes~Staguhn\altaffilmark{8,10}, 
J\"{u}rgen~Stutzki\altaffilmark{8}, 
Christopher~K.~Walker\altaffilmark{11}, 
Robert~W.~Wilson\altaffilmark{1,7}, 
Gregory~A.~Wright\altaffilmark{7}, 
Xiaolei~Zhang\altaffilmark{1,10},
Peter~Zimmermann\altaffilmark{12},
and 
R\"{u}diger~Zimmermann\altaffilmark{12}}
\affil{The Center for Astrophysical Research in Antarctica}

\altaffiltext{1}{Harvard-Smithsonian Center for Astrophysics; 60 Garden St.; 
Cambridge, MA 02138 ({\tt aas, adair, cmartin@cfa.harvard.edu}).}
\altaffiltext{2}{CASA; University of Colorado; Boulder, CO 80309.} 
\altaffiltext{3}{Boston University; 725 Commonwealth Ave.; Boston, MA 02215.}
\altaffiltext{4}{Caltech Submillimeter Observatory; 
111 Nowelo St.; Hilo, HI 96720. }
\altaffiltext{5}{University of Illinois; 1002 W. Green St.;
Urbana, IL 61801.}
\altaffiltext{6}{Radio Astronomy Lab; 
University of California; Berkeley, CA 94720 ({\tt greg@astron.berkeley.edu}).}
\altaffiltext{7}{Bell Laboratories; 791 Holmdel-Keyport Rd.; Holmdel, NJ 
07733.}
\altaffiltext{8}{I. Physikalisches Institut, Universit\"{a}t zu K\"{o}ln;
Z\"{u}lpicher Stra{\ss}e 77; D-50937 K\"{o}ln, Germany.}
\altaffiltext{9}{California Institute of Technology; Pasadena, CA 91125.}
\altaffiltext{10}{NASA GSFC; Code 685; Greenbelt, MD 20771 ({\tt
staguhn, zhang@stars.gsfc.nasa.gov}). }
\altaffiltext{11}{Steward Observatory; 933 N. Cherry Ave.; University of 
Arizona; Tucson, AZ 85721.}
\altaffiltext{12}{Radiometer Physics GmbH; Birkenmaarstra{\ss}e 10; 53340 
Meckenheim,
Germany.}

\begin{abstract}

AST/RO, a 1.7 m diameter telescope for astronomy 
and aeronomy studies at wavelengths between 
200 and 2000 $\mu \rm m$, 
was installed at South Pole during the 1994-95 
Austral summer. The telescope
operates continuously through the Austral winter,    
and is being used primarily for spectroscopic studies of neutral atomic
carbon and carbon monoxide in the interstellar medium of the Milky Way 
and the Magellanic Clouds.
The South Pole environment is unique among observatory
sites for unusually low wind speeds,
low absolute humidity, and the consistent clarity of the submillimeter sky.
Especially significant are the exceptionally low values of
sky noise found at this site, a result of the small water vapor content
of the atmosphere.
Four heterodyne
receivers, an array receiver, 
three acousto-optical spectrometers, and an array spectrometer
are currently installed.   A Fabry-Perot spectrometer using a 
bolometric array and a Terahertz receiver are in development.
Telescope pointing, focus, and calibration methods as well as 
the unique working environment and logistical requirements 
of South Pole are described.

\end{abstract}


\section{Introduction}

Most submillimeter-wave radiation from astronomical
sources is absorbed by irregular concentrations of 
atmospheric water vapor
before it reaches the Earth's surface;
ground-based submillimeter-wave observations are beset
by nearly opaque, variable skies.
Astronomers have therefore sought high, arid sites for 
new submillimeter telescopes,
to find better transparency and reduced sky noise. 
Among the most promising sites for submillimeter-wave astronomy
is South Pole, an exceptionally dry and cold site 
which has unique logistical opportunities and challenges.

\begin{figure}
\plotone{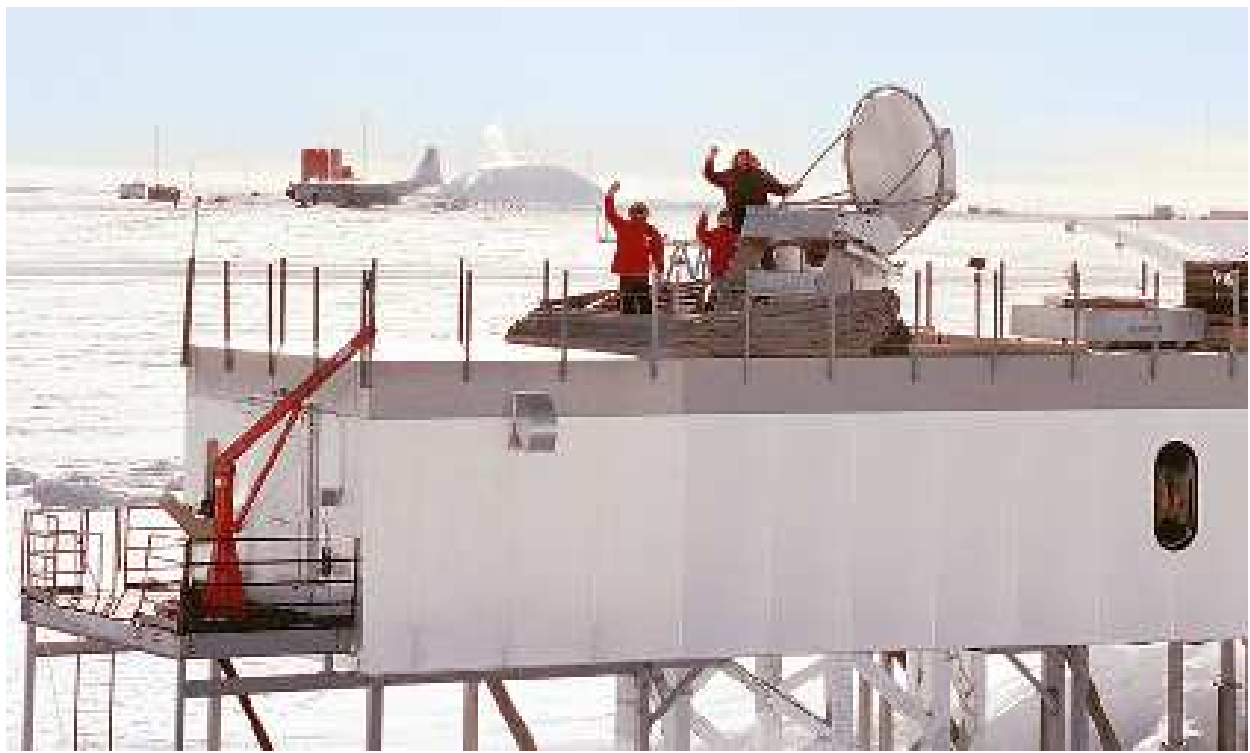}
\caption[AST/RO at South Pole]
{
{\bf AST/RO at South Pole.\ \ } The Antarctic Submillimeter Telescope
and Remote Observatory atop its building at South
Pole in February 1997.  Standing next to the telescope are 
G. Wright of Bell Labs, and X. Zhang and A. Stark of SAO.
The telescope rests on a steel support tower which is structurally
isolated from the building, and can be covered by a fold-off canopy made
of canvas and aluminum tubing.
An LC130
cargo aircraft is parked on the skiway.  The South Pole itself is
slightly below and to the right of the tail of the airplane. 
(photo credit: A. Lane)
\label{fig:astro} 
}
\end{figure}

Summer-only submillimeter observations from
South Pole were begun by \citet{pajot89}
and \citet{dragovan90}.
A year-round observatory has been established
in the past decade
by the Center for Astrophysical 
Research in Antarctica (CARA),
an NSF Science and Technology Center.  
CARA has fielded several major telescope facilities:  
AST/RO (the Antarctic
Submillimeter Telescope and Remote Observatory, a 1.7-m telescope  
(see Figure~\ref{fig:astro}),  
Python and Viper (Cosmic Microwave Background experiments), 
DASI (the Degree-Angular Scale Interferometer), and SPIREX (the South Pole
Infrared Explorer, a 60-cm telescope, now decommissioned).
These facilities are conducting site characterization and astronomical 
investigations from millimeter wavelengths to the near-infrared 
\citep[see also URL {\tt http://astro.uchicago.edu/cara}]{landsberg98}. 
AST/RO was installed at South Pole in Austral 
summer 1994-1995 \citep{lane96a}, 
and was the first submillimeter-wave
telescope to operate on the Antarctic Plateau in winter.
AST/RO is designed for
astronomy and aeronomy with heterodyne and bolometric detectors
at wavelengths between 3 mm and $200 \micron$. An up-to-date list of
scientific publications and technical memoranda from AST/RO may be found
at the observatory website at 
{\tt http://cfa-www.harvard.edu/}$\sim${\tt adair/AST}\_{\tt RO}.
Observing time on 
AST/RO is now open to proposals from the worldwide astronomical 
community.

This paper describes observing operations at AST/RO.  A summary
of logistical support requirements and 
measured site characteristics is given in \S 2.  
The instrument and
observatory facilities are described in \S 3 and in \citet{stark97a}.
Telescope pointing, calibration, and observing methods are described in \S 4.

\begin{table}
\label{table:environment}
\vspace{0.01in}
\caption{Physical Environment of the AST/RO Telescope}
\begin{center}
\begin{tabular}{l r}
Altitude& 2847 m\\
Longitude& W $45^o 53'$\\
Latitude& S $89^o 59' 40''$\\
Average Pressure\tablenotemark{a}& 680 mb\\
Minimum Temperature\tablenotemark{a}& $-82$ C \\
Maximum Temperature\tablenotemark{a}& $-14$ C\\
Average Temperature\tablenotemark{a}& $-49$ C \\
Average 24-hour Temperature variation\tablenotemark{b}& $6.1$ C \\
Average Wind Speed\tablenotemark{a}& $5.8\, \rm  m\, s^{-1}$\\
Maximum Wind Speed\tablenotemark{a}& $24\, \rm  m\, s^{-1}$\\
Annual Average Cloud Cover\\
\quad clear&31\%\\
\quad scattered&27\%\\
Rainfall\tablenotemark{a}&  0 \\
Median Water Vapor Column in summer\tablenotemark{c}&  0.47 mm PWV\\
Median Water Vapor Column in winter\tablenotemark{c}&  0.25 mm PWV\\
Best 5\% Water Vapor Column in winter\tablenotemark{c}&  0.10 mm PWV\\
Median Observed Zenith Transmission at $609 \mu \rm m$ &\\
\quad Wavelength in winter\tablenotemark{c} & 0.50\\
Median Calculated Zenith Transmission at $350 \mu \rm m$ &\\
\quad Wavelength in winter\tablenotemark{d} & 0.58\\
\end{tabular}
\end{center}
\tablenotetext{\rm a}{Schwerdtfeger 1984}
\tablenotetext{\rm b}{in 1992}
\tablenotetext{\rm c}{Lane 1998}
\tablenotetext{\rm d}{Bally 1989}
\end{table}

\section{Site Characteristics}

\subsection{Logistics}
The AST/RO telescope is located in the {\em Dark Sector} of 
the United States 
National Science Foundation Amundsen-Scott South Pole Station.
The station provides logistical support for the observatory: room and board
for on-site 
scientific staff, electrical power, network and telephone connections,
heavy equipment support, and cargo and personnel transport.
The station powerplant provides about 25 kW of power to 
the AST/RO building out of a total generating capacity of about 490 kW.  
South Pole has been continuously populated since the first station was
built in November 1956.  The current station 
was built in 1975, and new structures
have been added in subsequent years 
to bring the housing capacity to 210 people in
Austral summer and 45 in the Austral winter.  
New station facilities are under construction and are expected to be
operational by 2005.  These include living quarters for a winter-over
staff of 50, a new powerplant with greater generating capacity, and
a new laboratory building.

Heavy equipment at South Pole Station 
includes cranes, forklifts, and bulldozers;
these can be requisitioned for scientific use as needed.
The station is supplied by over 200 
flights each year of LC130 ski-equipped cargo
aircraft.  Annual cargo capacity is about 3500 tons.
Aircraft flights are scheduled 
from late October to early-February so that the station
is inaccessible for as 
long as nine months of the year.  This long {\em winter-over} 
period is central to all 
logistical planning for Polar operations.  

The SIS receivers used on AST/RO each require about 2 liters of liquid
helium per day.  As of mid-2000, 
total usage of liquid helium at South Pole averages 
30 liters per day for all experiments. One or two helium-filled
weather balloons are launched each day. 
There is an ongoing loss of helium from
the station, as it is used for a variety of experiments. 
The National Science Foundation and its support contractors  
must supply helium to South Pole, and
the most efficient way to transport and supply helium is in liquid form.
Before the winter-over period, one or more large (4000 to 12000 liter) storage
dewars are brought to
South Pole for winter use; some years this supply lasts the
entire winter, but in 1996 
and 2000 it did not.  In December 1998, no helium was 
available during a period scheduled for engineering tests, resulting
in instrumental problems for AST/RO during the subsequent winter. 
The supply of liquid helium has been a chronic problem
for AST/RO and for South Pole astronomy, but improved facilities in the new
station should substantially improve its reliability.

Internet and telephone service to South Pole is
provided by a combination of two low-bandwidth satellites, LES-9 and GOES-3,
and the high-bandwidth (3 Mbps) NASA 
Tracking and Data Relay Satellite System TDRS-F1.
These satellites are geosynchronous but not geostationary, since
their orbits are inclined.
Geostationary satellites are always below the horizon and cannot be used.
Internet service is intermittent through each 24-hour period because 
each satellite is visible only during the southernmost part
of its orbit; the combination of the three satellites provides an Internet
connection for approximately 12 hours within the period 1--16 hr Greenwich LST.  
The TDRS link helps provide a
store-and-forward automatic transfer service for large computer files.
The total data communications capability is about 5 Gbytes per day.
AST/RO typically generates 1--2 Mbytes per day.
Additional voice communications are provided by a fourth satellite, ATS-3, 
and high frequency radio.

On AST/RO, all engineering operations for equipment installation and
maintenance are tied to the annual cycle of physical access to the instrument. 
Plans and schedules are made in March and April for each year's deployment 
to South Pole:
personnel on-site, tasks to be completed, and the tools and equipment needed.
All equipment must be ready for shipment by the end of September.  Orders 
for new equipment should be 
complete by June and new equipment should be tested and
ready to ship by August.  
For quick repairs and upgrades, it is possible to send
equipment between South Pole and anywhere serviced by 
commercial express delivery in about
five days during the Austral summer season.

AST/RO group members deploy to South Pole in groups of 
two to six
people throughout the Austral summer season, carry out their 
planned tasks as well as circumstances allow, and return, after stays
ranging from 2 weeks to 3 months.   
Each year there is an AST/RO {\em winter-over scientist}, a single person
who remains with the telescope for one year.   The winter-over scientist
position is designed to last three years: one year of preparation and
training, one year at South Pole with the telescope, and one year 
after the winter-over year to reduce data and prepare scientific results.
If there are no instrumental difficulties, the winter-over scientist 
controls telescope observations 
through the automated control program {\em OBS} (see \S 4),
carries out routine pointing and calibration tasks, tunes the
receivers, and fills the liquid
helium dewars.  If instrumental difficulties develop, the winter-over scientist
carries out repairs in consultation with 
AST/RO staff back at their home institutions
and with the help of other winter-over staff at South Pole.

AST/RO is located on the roof of a dedicated support building across the 
aircraft skiway in the {\em Dark Sector}, a grouping of observatory buildings
in an area designated to have low radio emissions and light pollution.
The AST/RO building is a single story, $4 {\rm m} \times 20 {\rm m}$,
and is elevated $3 {\rm m}$ above the surface on steel columns to reduce 
snow drifts.  The interior is partitioned into six rooms, including laboratory
and computer space, storage areas, a telescope control room, and a 
Coud\'e room containing the receivers on a large optical table directly 
under the telescope.

\subsection{Instrument Reliability}

AST/RO is a prototype, the first submillimeter-wave telescope to
operate year-round on the Antarctic plateau.   As such, its operation
has been an experiment, minimally staffed and supported, 
intended to demonstrate feasibility and
to identify areas of difficulty.  A unique challenge of South Pole
operations is the lack of transport for personnel and equipment
during the nine month winter-over period.  Spare or replacement parts 
for most of the critical system assemblies
have been obtained, shipped to South Pole station, and stored in the AST/RO
building.
In a typical year, failure occurs in two or three system subassemblies, 
such as a drive system power supply or a submillimeter-wave local 
oscillator chain.  Usually a repair or work-around is effected by
the winter-over scientist.
There are, however, single points of failure which can cause 
the cessation of observatory operations until the end of the winter.
Between 1995 and 2000, observations or engineering tests were
planned for a total of 54 months, of which about half were
successful.  The most important cause of
telescope downtime has been incapacitation
of the single winter-over
scientist, resulting in 14 months of lost time.
Failure of the liquid helium supply was
responsible for a further 11 months of lost time.
In future, both of these causes for failure will be substantially
reduced.  Starting in 2002, it is planned that AST/RO will field 
two winter-over
scientists.  Beginning in 2003, the completion of a new liquid
helium supply facility as part of the new South Pole Station
Modernization plan will eliminate single points of failure
for the liquid helium supply.

\subsection{Site Testing}

The sky is opaque to submillimeter wavelengths at most observatory sites.
Submillimeter astronomy can only be pursued from 
dry, frigid sites, where the atmosphere contains 
less than 1 mm of precipitable
water vapor (PWV).  Water vapor is usually the dominant source of opacity,
but thousands of other molecular lines 
contribute \citep{waters76,bally89} a
{\em dry air} component to the opacity.  
\cite{chamberlin95},
\cite{chamberlin97}, and 
\cite{chamberlin00} showed that the dry air opacity
is relatively more important at the South Pole than
at other sites.  
Dry air opacity is less variable than the opacity
caused by water vapor, and therefore causes less
{\em sky noise}.

\begin{figure}
\plotone{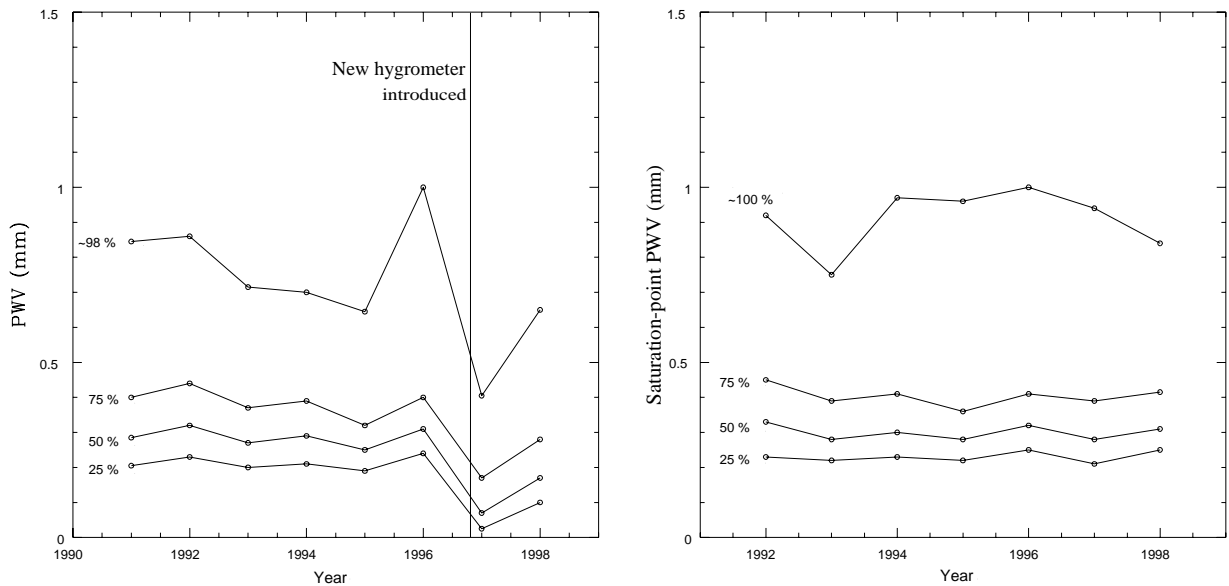}
\caption[Balloon measurements of PWV at South Pole]
{
{\bf Balloon measurements of PWV at South Pole.\ \ }
(left) Quartiles of precipitable water vapor (PWV) distribution 
in winter (day-of-year 100--300) from 1991 through 1998,
calculated from balloon-borne radiosonde measurements.
The radiosonde type was changed in early 1997, and the subsequent
calibration has been spurious.
This new radiosonde, the AIR model 5a, has been used in
measurements at other sites, in particular the peaks
surrounding Chajnantor plateau \citep{giovanelli99}; the
1997 South Pole measurements make possible a direct comparison. 
(right) Quartiles of saturation-point PWV distribution 
in winter (day-of-year 100-300) from 1992 through 1998,
calculated from balloon-borne pressure and temperature measurements.
Saturation-point PWV is calculated by assuming 100\% water
vapor saturation for a column of air with a measured
temperature and pressure profile.  Since temperature and pressure
sensor calibration is more reliable than hygrometer calibration,
this value gives reproducible results.
If hygrometer measurements from 1991-1996 can be trusted, the
winter South Pole atmosphere is, on average, 90\% saturated.
\label{fig:hygrometer}}
\end{figure}
\begin{figure}
\plotone{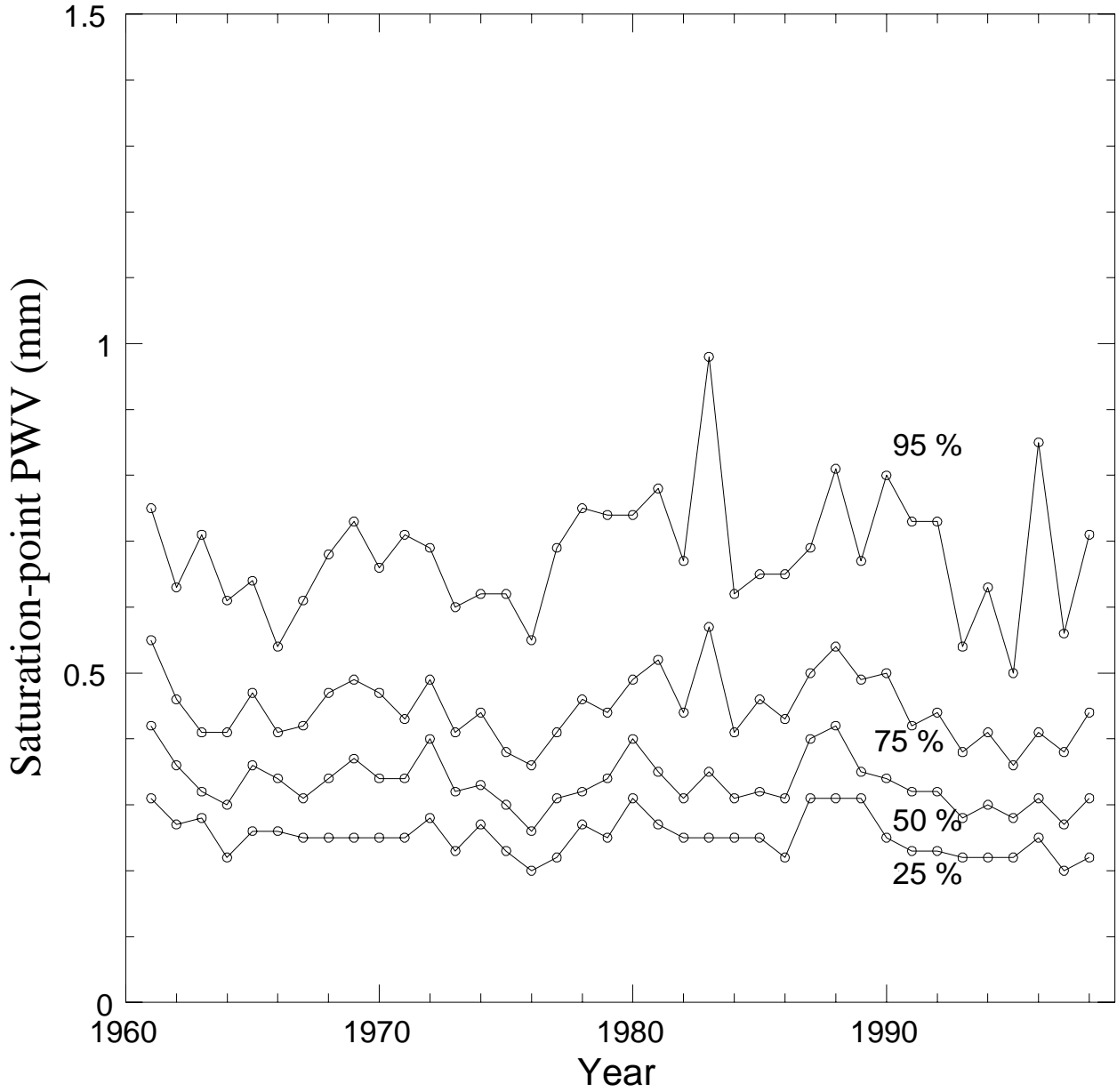}
\caption[Saturation-point PWV at South Pole, 1961-1998]
{
{\bf Saturation-point PWV at South Pole, 1961-1998, adapted 
from Chamberlin (2000).\ \ }
Quartiles of saturation-point PWV distribution during the winter
(day-of-year 100-300) for 1961 through 1998,
calculated from balloon-borne pressure and temperature measurements.
This figure illustrates the long-term stability of the South Pole
climate.
\label{fig:satpoint5998}
} 
\end{figure}
\begin{figure}
\plotone{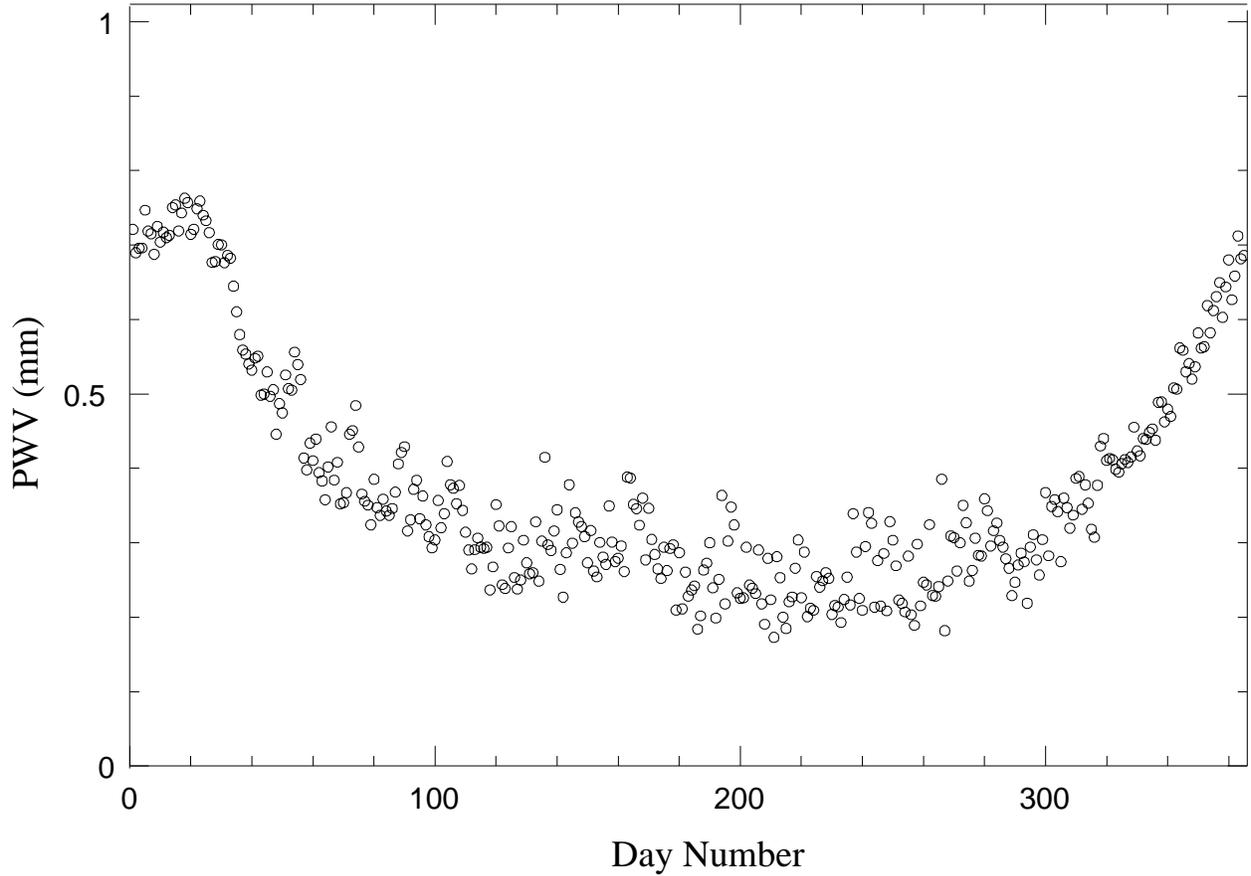}
\caption[Average PWV at South Pole by day of year, 1961-1999]
{
{\bf Average PWV at South Pole by day of year, 1961-1999, adapted 
from Chamberlin (2000).\ \ }
The PWV in millimeters for each day of the year
between 1961 and 1999 is
averaged and plotted as a function of day of year.  This
plot shows the average seasonal trend, where the average 
water vapor content of the atmosphere declines from 
February through September.
\label{fig:PWVaveyear}
} 
\end{figure}

Physical parameters of the South Pole site of AST/RO are given in Table 1.
The South Pole meteorology office has used 
balloon-borne radiosondes to measure
profiles above the South Pole of temperature, pressure, and water vapor
at least once a day for several decades \citep{schwerdtfeger}.
These have typically shown atmospheric water vapor values 
about 90\% of saturation for air coexisting
with the ice phase at the observed temperature and pressure.
The precipitable water vapor (PWV) values consistent with saturation are, 
however, extremely low because the 
air is dessicated by the frigid temperatures. 
At the South Pole's average annual temperature of $-49$ C, the partial
pressure of saturated water vapor is only 1.2\% of what 
it is at 0 C \citep{goff46}.
Judging by other measures of PWV such as LIDAR and mid-infrared spectroscopy,
the calibration of the hygrometers used on balloon sondes was accurate
between 1991 and 1996.  On February 22, 1997, the balloon radiosonde
type was changed from an A.I.R. Model 4a to the A. I. R. Model 5a, and
the average PWV values indicated by the new radiosondes dropped by
70\% (see Figure~\ref{fig:hygrometer});
these new values appear to be spuriously low.  A firm upper limit
to the PWV can be set by calculating what the PWV would be if the
column of air were 100\% saturated with water vapor at the observed temperature
and pressure, the {\em saturation-point PWV}.  
Since the temperature and pressure measurements from balloon
sondes are accurate, and since the atmosphere cannot be significantly
supersaturated, the saturation-point PWV is 
a reliable upper limit to the true PWV.  Values of the saturation-point PWV
for a 38 year period are shown in Figure~\ref{fig:satpoint5998}
\citep{chamberlin00}.  
Figure~\ref{fig:PWVaveyear} shows the PWV for each day of the year
averaged over all the years in this data set
to show the average seasonal variation in the water vapor
content of the atmosphere.
{\em PWV values
at South Pole are small, stable, and well-understood.}  

\begin{figure}
\plotone{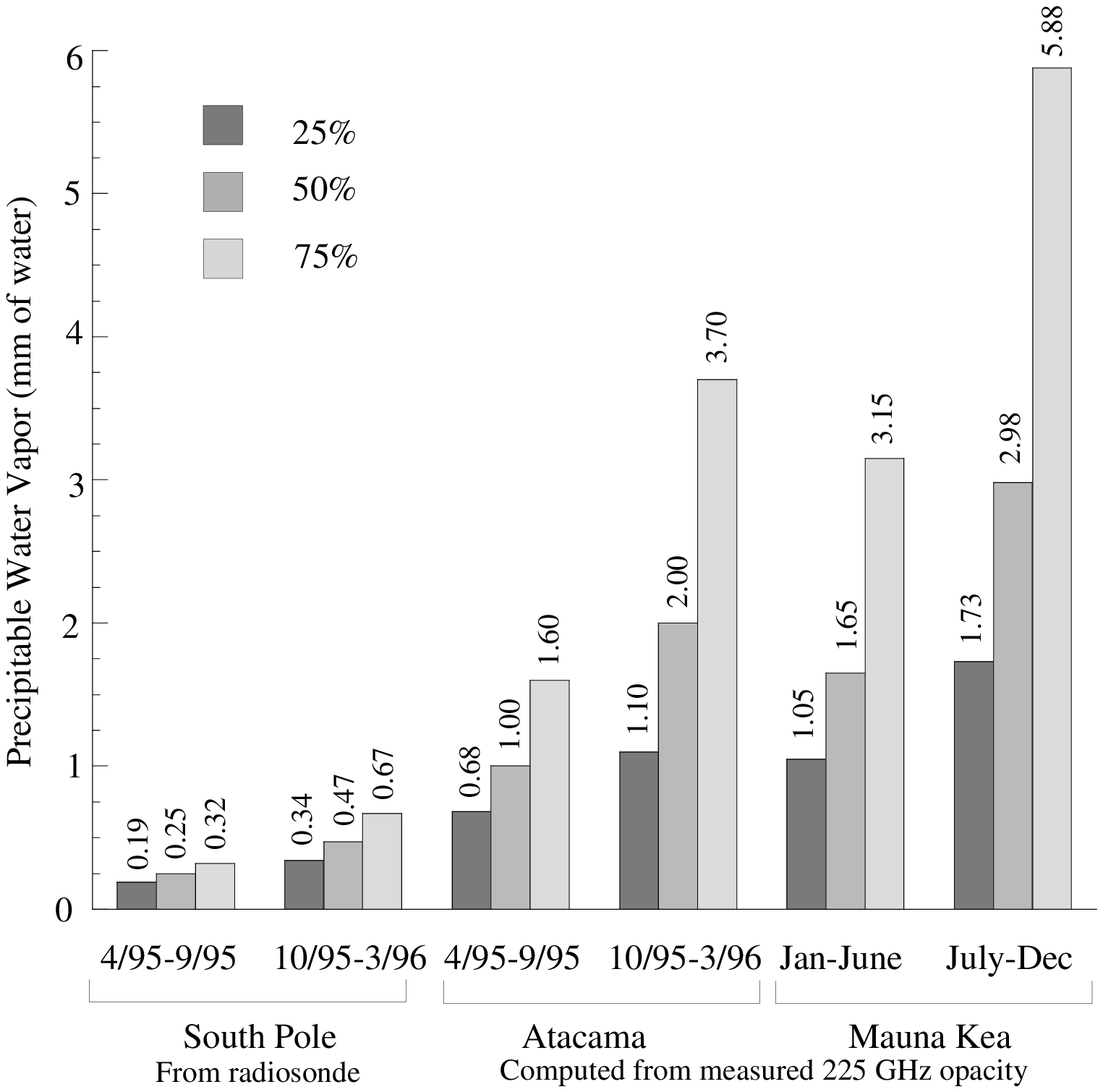} 
\caption[Quartiles of Precipitable Water Vapor at three sites.]
{{\bf Quartiles of Precipitable Water 
Vapor at three sites, from Lane (1998).\ \ }
South Pole is a considerably drier 
site than Mauna Kea or Atacama (Chajnantor).
During
the wettest quartile at South Pole, the total precipitable water vapor
is lower than during the driest quartile at either Mauna Kea or Atacama.
\label{fig:PWVbargraph} 
}
\end{figure}

Quartile values of the distribution of PWV with time are plotted in 
Figure~\ref{fig:PWVbargraph} \citep[from][]{lane98}, where 
they are compared with 
corresponding values for Mauna Kea and for 
the proposed ALMA site at Chajnantor.
The relation between PWV and measured opacity 
at 225~GHz \citep[S. Foster, private communication]{hogg92,masson94}
was used to derive the PWV values for Mauna Kea and Chajnantor.
Recent Fourier Transform Spectrometer results from Mauna Kea \citep{pardo01a} 
show that this PWV-$\tau_{\mathrm{225}}$ relation may overestimate PWV
by about 12\%.
The data are separated into the best 6-month period and 
the remainder of the year. 
Of the three sites, South Pole has by far the
lowest PWV, during Austral summer as well as winter. 

Millimeter and 
sub\-millimeter-wave atmospheric
opacity at South Pole has been measured using skydip techniques. 
\citet{chamberlin97} made
over 1100 skydip observations at 492~GHz (609 \micron ) with AST/RO during the
1995 observing season. 
Even though this frequency is near a strong oxygen line, the opacity was
below 0.70 half of the time during the Austral winter and reached values as
low as 0.34, better than ever measured at any ground-based site. The
stability was also remarkably good: the opacity remained below 1.0 for weeks
at a time. 
The 225~GHz (1.33 mm) skydip data for South Pole were
obtained during 1992 \citep{chamberlin94,chamberlin95} using a standard NRAO
tipping radiometer similar to the ones used to measure the 225 GHz zenith
opacities at Mauna Kea and  Chajnantor,
and the results are summarized by \citet{chamberlin97} and \citet{lane98}.
The tight linear relation between 225 GHz skydip data and balloon sonde PWV 
measurements is discussed by \citet{chamberlin95}.

\begin{figure}
\epsscale{0.6666}
\plotone{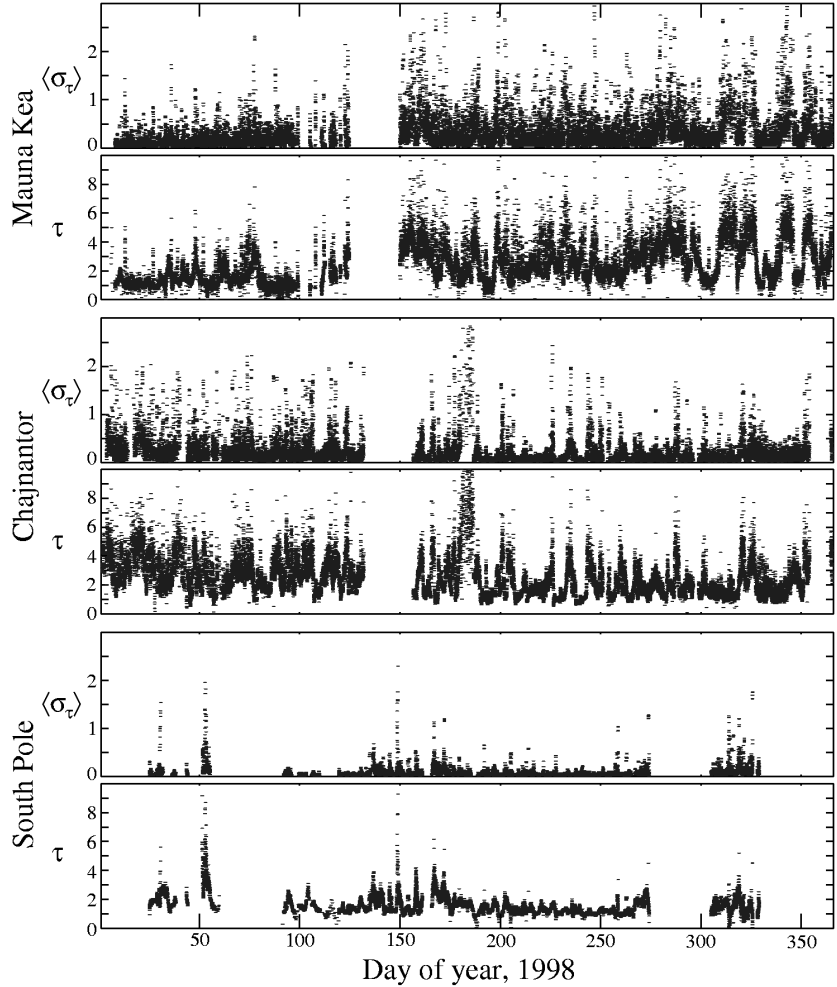}
\caption[Sky noise and opacity 
measurements at 350 \micron \ from three sites]
{{\bf Sky noise and opacity measurements 
at 350 \micron \ from three sites.\ \ } 
These plots show data from identical NRAO-CMU 350 \micron \ broadband tippers 
located at Mauna Kea, the ALMA site at Chajnantor, and South Pole during 1998.  The upper
plot of each pair shows $\langle \sigma_\tau \rangle$, the rms deviation
in the opacity $\tau$ during a one-hour period---a measure
of sky noise on large scales; the lower plot of each
pair shows $\tau$, the broadband 350 \micron \ opacity.
The first 100 days of 1998 on Mauna Kea were exceptionally good for that site.  
During the best weather at
South Pole, $\langle \sigma_\tau \rangle$ was dominated by 
detector noise rather than sky noise.
Data courtesy of S. Radford and J. Peterson.
}
\label{fig:threesites}
\end{figure}

    From early 1998, the 350$\mu$m (850 GHz) band 
has been continuously monitored 
at Mauna Kea, Chajnantor, and South Pole by identical tipper instruments
developed by S. Radford of NRAO and J. Peterson of Carnegie-Mellon U. and
CARA.  Results from South Pole are compared to Chajnantor and Mauna Kea
in Figure~\ref{fig:threesites}.
These instruments measure a broad band that includes the center of the
350 \micron \, window as well as more opaque nearby wavelengths.  Comparison
of the $\tau$ values measured by these instruments is tightly correlated with
occasional narrow-band skydip measurements made within this band 
by the CSO and AST/RO; the narrow band $\tau$ values are about a
factor of two smaller than those output by the broadband instrument.
{\em The 350\micron \ opacity at South Pole is consistently better
than at Mauna Kea or Chajnantor.}  

 
\begin{figure}
\epsscale{0.5}
\plotone{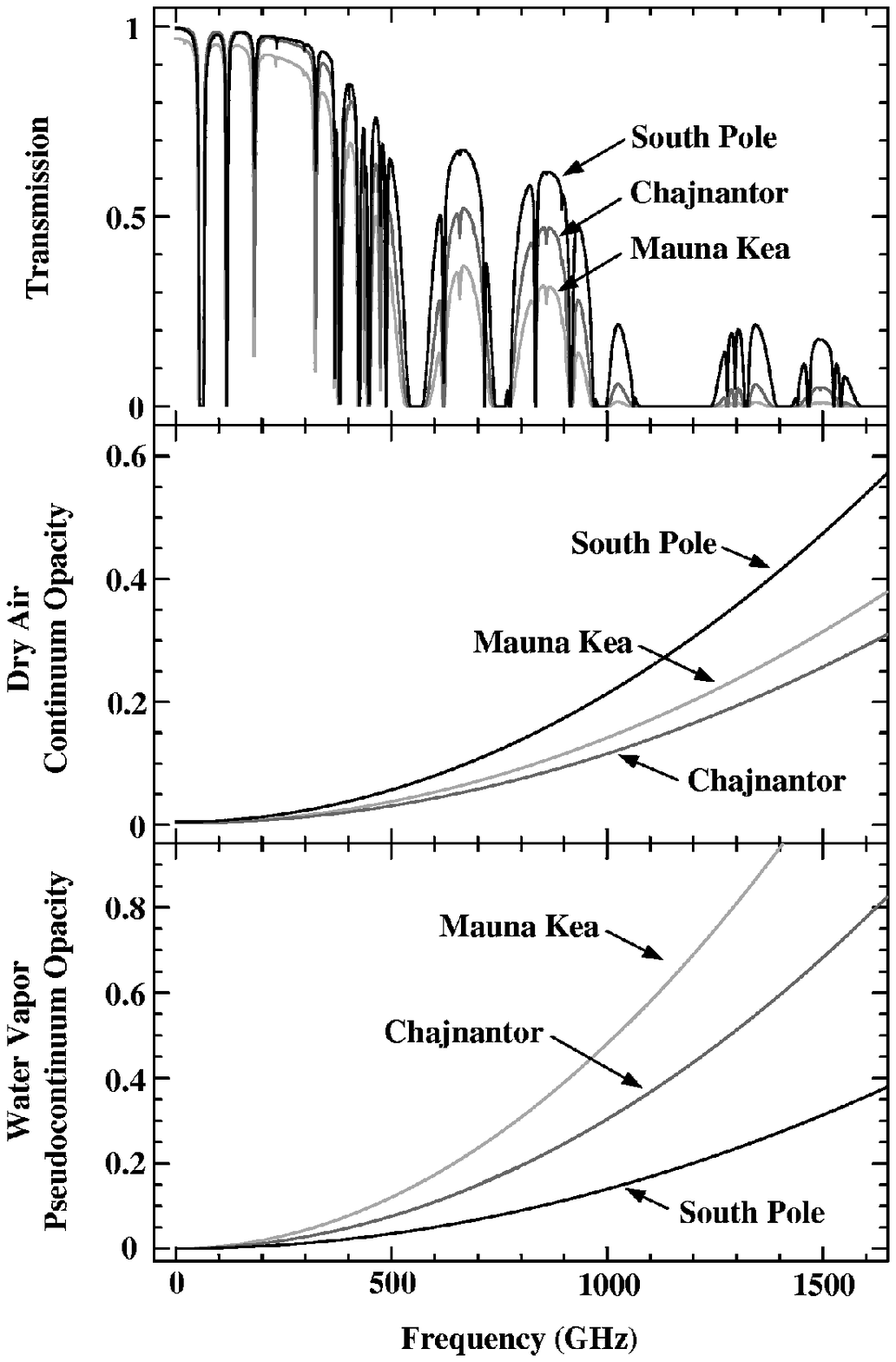}
\caption[Calculated atmospheric transmittance at three sites]
{
{\bf Calculated atmospheric transmittance at three sites.\ \ }
The upper plot is atmospheric transmittance at zenith
calculated by J. R. Pardo using the ATM model \citep{pardo01b}.
The model uses PWV values of 0.2 mm for South Pole, 0.6 mm
for Chajnantor and 0.9 mm for Mauna Kea, corresponding to
the $25^{\mathrm{th}}$ percentile winter values at each site.
Note that at low frequencies, the Chajnantor curve converges with the
South Pole curve, an indication that 225 GHz opacity is not a simple
predictor of submillimeter wave opacity.
The middle and lower plots show calculated values of dry air
continuum opacity and water vapor pseudocontinuum opacity for
the three sites.  Note that unlike the other sites, the opacity 
at South Pole is
dominated by dry air rather than water vapor.
}
\label{fig:pardo} 
\end{figure}

The simulations shown in Figure \ref{fig:pardo} 
were made with the ``Atmospheric Transmission 
at Microwaves" (ATM) model \citep{pardo01b}.
They include all water and oxygen lines up to 10 THz 
and account for the non-resonant excess of water vapor absorption and 
collision-induced absorption of the dry atmosphere according to the results
of \citet{pardo01a}.
For the comparison we have 
plotted typical transmission profiles for the three sites 
calculated for winter time 25\% PWV quartiles. 
For South Pole, we use 0.19 mm PWV
\citep{lane98}.  For Mauna Kea and 
Chajnantor we use the values 0.9 and 0.6 mm, respectively. 
The collision-induced 
non-resonant absorption of the dry atmosphere is higher at South Pole 
due to the higher ground level pressure compared with the other two sites. 
The much smaller PWV levels of South Pole, however, make its total zenith
submillimeter-wave transmission considerably better than Chajnantor
and Mauna Kea.

The success or failure of various submillimeter-wave observational 
techniques depends critically on atmospheric opacity.  
Just as it makes no sense to carry out visual-wavelength photometry
in cloudy weather, there is an atmospheric opacity above which any
particular submillimeter-wave observational technique will fail to
give usable results.  This threshold depends on the details of
the particular technique and its sensitivity to the spectrum of 
atmospheric noise, but for many techniques the threshold 
lies near $\tau \sim 1$.  Weather
conditions above the threshold cannot be compared to those below
the threshold by simple Gaussian noise analysis: ten days where
$\tau \approx 1.5$ can never 
be the equivalent of one day where $\tau \approx 0.5$.
For deep background experiments, it is important to choose the best possible
site.

{\em Sky noise} refers to fluctuations 
in total power or phase of a detector caused by
variations in atmospheric emissivity and path length on timescales of order one second.  
Sky noise causes systematic errors in the measurement of astronomical
sources.  In an instrument that 
is well-designed, meaning that it has no intrinsic
systematic noise \citep[see, for example,][]{kooi00}, 
the sky noise will determine the minimum flux
that can be observed, the flux below which the instrument will
no longer ``integrate down".  This flux limit is proportional to
the power in the sky noise spectral energy distribution at the 
switching frequency of the observing equipment.  
\cite{lay99} show analytically how sky noise causes observational techniques
to fail: fluctuations in a component of the data due to sky noise integrates 
down more slowly than $t^{-1/2}$ 
and will come to dominate the error during long
observations.  Sky noise is a source of systematic noise
which is not within the control of the instrument designer, and 
the limits to measurement imposed by sky noise are  
often reached in ground-based submillimeter-wave
instrumentation.

Sky noise at South Pole is considerably smaller than at other sites at the
same opacity.  As discussed by \citet{chamberlin95}, 
\citet{chamberlin00} and shown in Figure \ref{fig:pardo}, 
the PWV at South Pole is
often so low that the opacity is dominated by the {\em dry air} component;
the dry air emissivity 
and phase error do not vary as strongly or rapidly as
the emissivity and phase error due to water vapor. 
The spectral energy density of sky noise is determined
by turbulence in the atmosphere and has a roughly similar spectral shape 
at all sites \citep{lay99}.  Measurement of 
spectral noise at one frequency can therefore
be extrapolated to other frequencies.

Figure~\ref{fig:threesites} gives a direct 
indication of sky noise at submillimeter
wavelengths at the largest timescales.  The value 
$\langle \sigma_\tau \rangle$ is the root-mean-square deviation
of opacity measurements made within an hour's time.
As a measure of sky noise, this value has two defects:
(1) during the best weather it is limited by detector noise within the
NRAO-CMU tipper (which uses room-temperature bolometers) rather than
sky noise and
(2) the $\sim 10^{-3} \, {\rm Hz}$ fluctuations it measures are at
much lower frequencies than
the switching frequencies used for astronomical observations.
Therefore, $\langle \sigma_\tau \rangle$ is an upper limit to sky noise at
very low frequencies;
Figure~\ref{fig:threesites} nevertheless gives a clear 
indication that the power 
in sky noise at
South Pole is often several times less than at Mauna Kea or Chajnantor.

Other instruments are sensitive to sky noise
at frequencies near $10 \, {\rm Hz}$ and can be used to give quantitative
results over more limited periods of time. 
Sky noise at South Pole has been measured in conjunction with cosmic
microwave background experiments on Python 
\citep{Alvarez95,Dragovan94,Ruhl95,Platt97,coble99}
and White Dish \citep{tucker93}.  Python,
with a $2.75 \deg$ throw, had ${\mathrm{ 1\, mK \, Hz^{-1/2}}}$ sky noise on 
a median summer day, whereas White Dish, which had a $0.5 \deg$ throw, was
much less affected by sky noise.  Extrapolating to 218 GHz and a $0.2 \deg$
throw, the median sky noise is estimated to be 
${150\, \mu\mathrm {K\, Hz^{-1/2}}}$ even in the Austral summer, 
lower by a factor of ten than the sky noise
observed during Sunyaev-Zel'dovich (S-Z) effect observations on Mauna Kea
\citep{holzapfel97a}.  

\cite{lay99} have compared 
the Python experiment at South Pole
with the Site Testing Interferometer 
at Chajnantor \citep{radford96,holdaway95}.
These are very different instruments, but the differences can
be bridged by fitting to a parametric model. \cite{lay99} have
developed an atmospheric model for sky noise with a Kolmogorov power
law with both three- and two-dimensional regimes, and have
applied it to data from Python and the Chajnantor Testing
Interferometer.  They find that the amplitude of the sky noise at
South Pole is 10 to 50 times less than that at Chajnantor. {\em Sky noise
at South Pole is significantly less than at other sites.}

The best observing conditions occur  only at high elevation angles,
and at South Pole this means that only the southernmost 3 steradians of
the celestial sphere are accessible with the South Pole's uniquely
low sky noise---but this portion of sky includes
millions of galaxies and cosmological sources, the Magellanic clouds, and 
most of the fourth quadrant of the Galaxy.
The strength of South Pole as a submillimeter site lies in
the low sky noise levels routinely obtainable for sources 
around the South Celestial Pole.
This is crucial for large-scale
observations of faint cosmological sources observed with
bolometric instruments, and for such
observations South Pole is unsurpassed.

\begin{figure}
\epsscale{0.66666}
\plotone{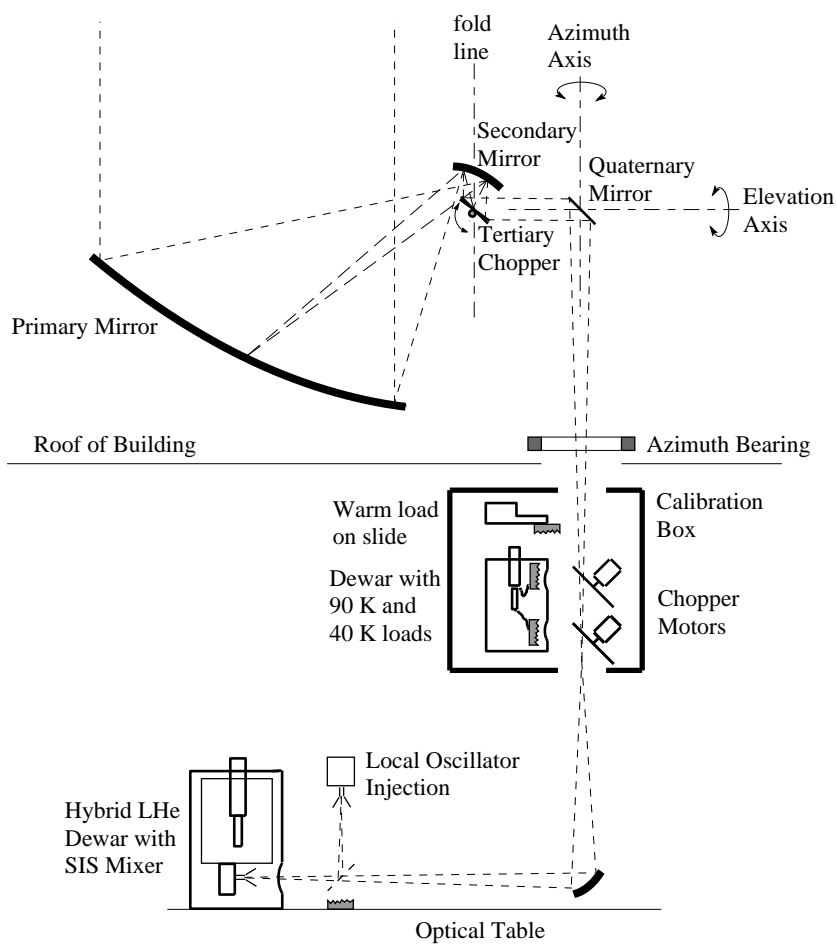}
\bigskip
\caption{ {\bf Schematic of the AST/RO optical system.}  For purposes
of representation, the beam path has been flattened and the reader
should imagine that the primary and secondary mirrors are rotated
by $90^{o}$ out of the plane of the page around the vertical 
``fold line".  
Note that rays diverging from a point on the primary mirror reconverge 
at the chopping tertiary mirror, since the tertiary is at the exit pupil 
of the instrument. 
The tertiary and quaternary mirrors are flat.
The calibration loads are in a dewar
to the side of the Coud\'e beam entering along the azimuth axis. The ambient
temperature load is on a linear actuator and can slide in front of the sky
port. Motorized chopper mirrors switch the receiver beam from the cooled loads
to the sky or to the ambient temperature load.
}
\label{fig:optics} 
\end{figure}

\section{Instrument and Capabilities}

The instrumental design of the
AST/RO telescope is described in \citet{stark97a}.  This
section describes the current suite of instrumentation and aspects
of the telescope optics which affect observations.

\subsection{Receivers}\label{S: receivers}
Currently, there are five heterodyne receivers mounted on an
optical table suspended from the telescope structure
in a spacious 
($5 \mathrm m \times 5 \mathrm m \times 3 \mathrm m$), warm Coud\'{e} room:
\begin{itemize}
\item{ a 230 GHz SIS receiver, 55--75 K
double-sideband 
(DSB) noise temperature;}
\item{ a 450--495 GHz SIS waveguide 
receiver, 200--400 K DSB \citep{walker92};}
\item{ a 450--495 GHz SIS quasi-optical 
receiver, 165--250 K DSB \citep{engargiola94, zmuidzinas92};}
\item{ a 800-820 GHz fixed-tuned 
SIS waveguide mixer receiver, 950--1500 K DSB \citep{honingh97};}
\item{ an array of four 800-820 GHz fixed-tuned 
SIS waveguide mixer receivers, 850--1500 K DSB 
\citep[][the PoleSTAR array]{groppi00}. }
\end{itemize}
A 1.5 THz receiver, TREND \citep{gerecht99} and an imaging Fabry-Perot 
interferometer, SPIFI \citep{swain98} are currently in
development.

\subsection{Spectrometers}\label{S: spectrometers}
There are four currently available
acousto-optical spectrometers (AOS):
\begin{itemize}
\item{two low resolution spectrometers 
(LRS) with a bandwidth of 1 GHz (bandpass 1.6--2.6 GHz); }
\item{an array AOS with four low 
resolution spectrometer channels with a bandwidth of 1 GHz (bandpass 1.6--2.6 GHz)
for the PoleSTAR array;
and }
\item{one high-resolution AOS (HRS) with 60 MHz bandwidth 
(bandpass 60--120 MHz).} 
\end{itemize}
The LRS are built with GaAs laser diodes at 785 nm
wavelength operating in single mode at about 25 mW power. The HRS
uses a 5 mW HeNe-laser at 632 nm wavelength. The stability of these
AOS are high, with  $> 200 \, \mathrm{s}$ Allan variance minimum time
for the two LRS and $> 300 \, \mathrm{s}$ for the 
HRS measured at the AST/RO
site under normal operating conditions. The designs of these AOS are
very similar to the AOS used at the KOSMA observatory
\citep{Schieder89}. 
All AOS are set up for nearly full Nyquist
sampling of the spectra, the pixel spacing of the LRS is 670 kHz at
1.1 MHz resolution bandwidth per pixel, while the HRS has 32 kHz pixel
spacing at 60 kHz resolution bandwidth. The fluctuation bandwidth of
the spectrometers, which is the effective bandwidth per channel appearing
in the radiometer equation,
is 50\% larger than the resolution bandwidth. 
This is what is expected in a diffraction-limited AOS design.
Aside from a single failure of the
HRS due to a faulty opto-coupler chip, all the AOS have been continuously
operating for 5 years without any problem.  Maintenance
requirements have amounted to a check and test once a year.
{\em This AOS design is reliable, stable, and accurate.}

\subsection{Optics}\label{S: optics}

All of the optics in AST/RO are offset
for high beam efficiency and avoidance of inadvertent
reflections and resonances.
Figure \ref{fig:optics} shows the optical arrangement in its Coud\'e form.
The primary 
reflector is made of carbon fiber and epoxy with a vacuum-sputtered aluminum
surface having a surface roughness of $6 \micron$ and an rms
figure of about $9 \micron$ \citep{stark95c}.
The Gregorian secondary is a prolate spheroid with its offset angle 
chosen using the method of \citet{dragone82}, so that the Gregorian
focus is equivalent to that of an on-axis telescope with the
same diameter and focal length.  The diffraction-limited 
field-of-view is $2^\circ$ in diameter at $\lambda 3 \rm mm $ and
$20'$ in diameter at $\lambda 200 \micron$.  The chopper
can make full use of this field-of-view, because it is located at the exit
pupil and so does not change the illumination pattern on the
primary while chopping.  

Note in Figure \ref{fig:optics} that rays diverging from a point on
the primary mirror reconverge at the tertiary mirror, since the tertiary
is at the exit pupil of the instrument. 
Optimizing the optics this way requires that the primary mirror
be cantilevered away from the 
elevation axis: this is accomplished with a truss of Invar rods which
hold the primary-to-secondary distance invariant with temperature.  
When the fourth mirror shown 
in Figure \ref{fig:optics} is removed, the telescope 
has a Nasmyth focus where the beam passes through an elevation bearing
which has a $0.2 \rm m$ diameter hole.  
Array detectors of various types can be used at this focus.

\section{Observing Considerations}

\begin{figure}
\plotone{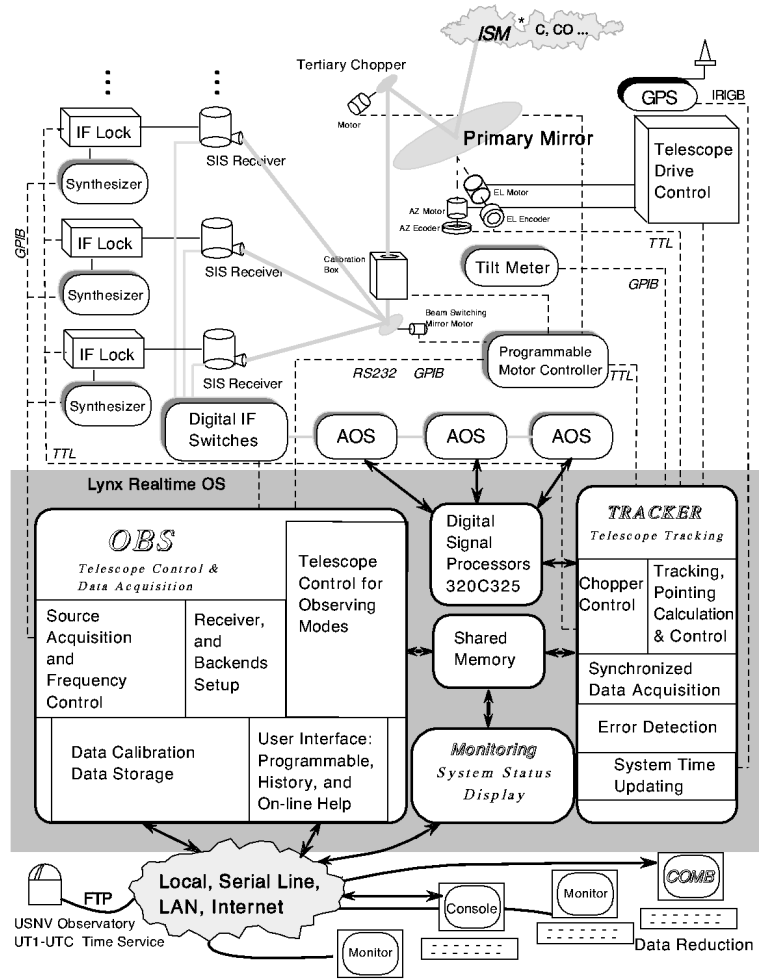}
\bigskip
\caption{ {\bf Schematic of the AST/RO control system
\citep{huang00}.}  
{This block diagram shows signal paths in the AST/RO
system.  Using a variety of interfaces and
protocols, the computer monitors and controls the hardware
subsystems to carry out the processes that constitute a
radioastronomical observation.
The data acquisition computer is shown as a shaded area.
}
}
\label{fig:maohai}
\end{figure}

Routine observations with AST/RO are automated,
meaning that if all is well the telescope acquires data
unattended for days at a time.   The observing control language for AST/RO
is called ``OBS" and was developed over a twenty-year period by R. W.
Wilson and collaborators in order to automate operations on the
Bell Laboratories 7m antenna.  
OBS-like languages are
described by \citet{stark93b,stark93c}.
OBS was modified for use on AST/RO by \citet{huang00}.
The signals controlled by OBS are shown in Figure \ref{fig:maohai}.
OBS commands and programs
can be submitted via the computer network, and remote control of observations
from sites beyond the South Pole station network is possible
during periods when communications satellites are
available.  For routine observations, winter-over
scientists normally monitor telescope operations from the
living and sleeping quarters at South Pole Station.   
This section describes aspects of the telescope and computer
interface which should be considered by the observer.
OBS commands for data acquisition and calibration are 
mentioned as appropriate.

\subsection{Focus}\label{S:focus}

\begin{figure}
\epsscale{0.66666}
\plotone{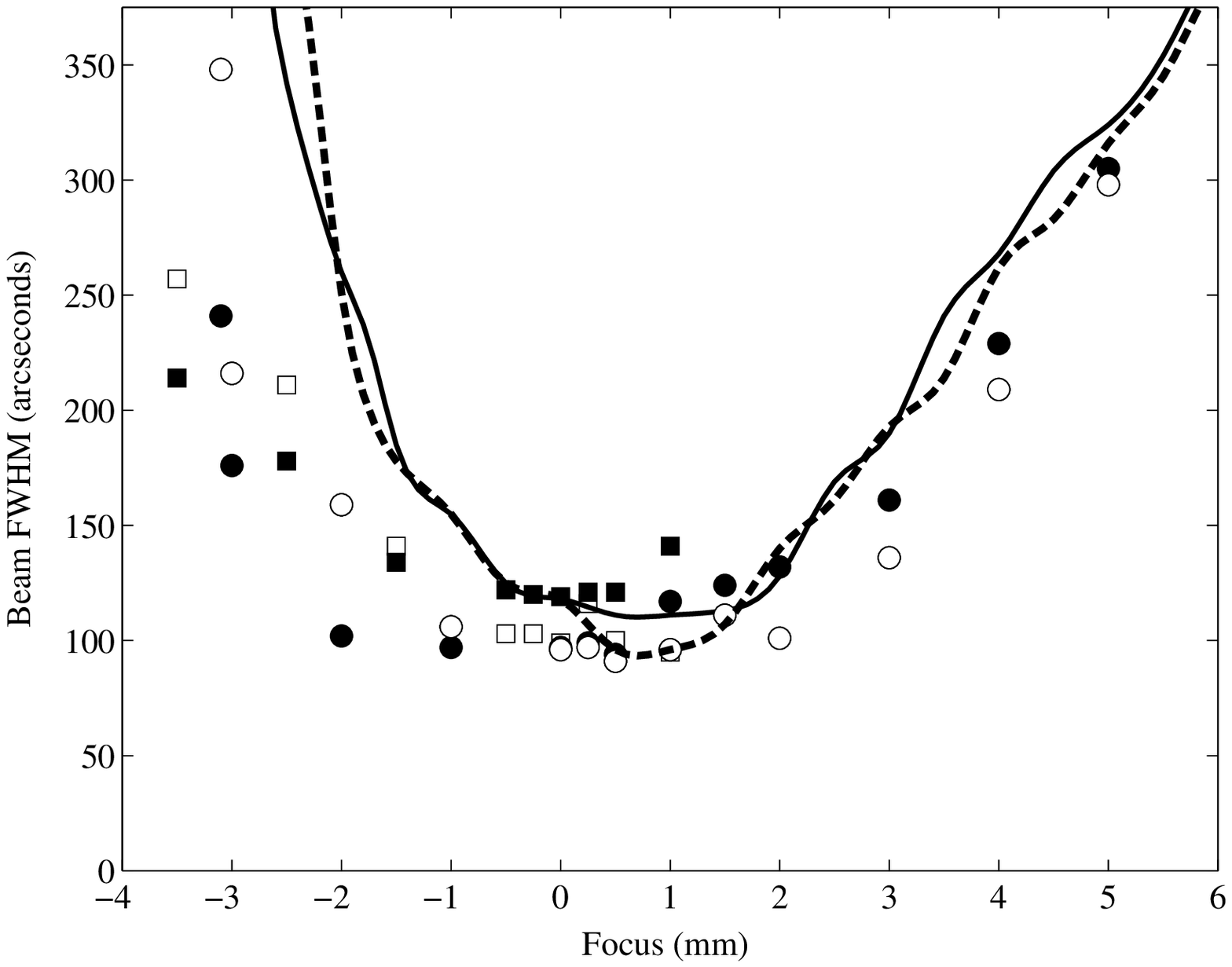}
\bigskip
\caption{{ \bf Theoretical and measured focus curves
for AST/RO.}  The horizontal (filled symbols, solid line) and
vertical (open symbols, dashed line) beam size plotted as
a function of focus adjustment.  Positive focus adjustment moves
the primary and secondary mirrors closer together.  Data plotted
as circles were obtained using the 460 GHz waveguide receiver;
data plotted as squares were obtained using the 492 GHz
quasi-optical receiver.  The solid and dashed lines represent
results calculated from a model of the AST/RO optical system,
assuming a 17.5 dB edge taper.
}
\label{fig:focus}
\end{figure}

The secondary mirror is mounted on a computer-controlled
positioning stage which allows the secondary mirror to be
translated in three dimensions: a focus translation along the
ray from the center of the primary to the center
of the secondary, translation parallel to the elevation axis,
and translation in a direction perpendicular to these two 
directions.  The angular alignment of the secondary is not
adjustable, but it does not need to be: as described in 
\citet{stark97a}, the primary mirror is aligned relative to
the secondary mirror by mechanical means.  This procedure
sets the angular alignment.
Figure \ref{fig:focus} shows the horizontal and vertical
beam size as a function of the focus adjustment, as measured
by scans of the Sun.  The data compare well with theoretical
expectation near the point of best focus where the beam has a
nearly gaussian shape; the deviations between theory and
measured points increase as the gaussian approximation fails
away from the point of best focus.

\subsection{Optical Pointing}

Telescope pointing is controlled by the main data acquisition
computer, shown schematically in Figure \ref{fig:maohai}.  
Input to the computer are time from the Global
Positioning Satellite receiver and encoder readouts from both
antenna axes.  Output from the computer are the commanded
velocities of the antenna axis motors.
The commanded velocity is calculated from the difference between
the encoder readouts and a {\em command position}, which is in
turn calculated from the time, the telescope position required
by the observing program,
and the {\em pointing model}.
The telescope drive system, as described in \citet{stark97a}, is
a mostly analog electronic servo mechanism
capable of driving the telescope so that the
encoder reading differs from the position commanded by the
computer by less than one second of arc for antenna velocities
near the sidereal rate.   The pointing model characterizes
the imperfections in the telescope construction, such as
gravitational sag, bearing misalignments, and manufacturing errors. 

\paragraph{Refraction.}
The pointing is corrected for refraction in 
the Earth's atmosphere by an approximate formula:
\begin{equation}
\Delta El (\mathrm{refraction}) = (n - 1) \, \mathrm{cot}( El ) ~~~,
\label{eq:refraction}
\end{equation}
where $n$ is the index of refraction,
\begin{equation}
N \equiv 10^6 \cdot (n - 1) \cong
77.6 \left [ {{P}\over{1 \, {\mathrm{mb}}}} \right ] \! \left [ {1 \, {\mathrm{K}}\over{T_{\mathrm{surf}}}} \right ]
- 6 \left [ {{P_{\mathrm{water}}}\over{1 \, {\mathrm{mb}}}} \right ] \!
\left [ {1 \, {\mathrm{K}}\over{T_{\mathrm{surf}}}} \right ]
+ \xi(\nu) \, 3.75 \times 10^5 \left [ {{P_{\mathrm{water}}}\over{1 \, {\mathrm{mb}}}} \right ] \!
\left [ {1 \, {\mathrm{K}}\over{T_{\mathrm{surf}}}} \right ]^2 ~,
\label{eq:index}
\end{equation}
$P$ is atmospheric pressure at the telescope, $P_{\mathrm{water}}$ is the partial pressure of atmospheric water,
and $T_{\mathrm{surf}}$ is the atmospheric temperature at the Earth's surface
\citep{bean62}.
The apparent elevation of sources is larger than it would be if the
atmosphere were absent.
At low frequencies, radio waves partially polarize atmospheric water vapor molecules,
increasing the refraction compared to 
visual wavelengths.
The third term in $N$,  
$3.75 \times 10^5 \left [ {{P_{\mathrm{water}}}/{1 \, {\mathrm{mb}}}} \right ] 
\left [ {1 \, {\mathrm{K}}/{T_{\mathrm{surf}}}} \right ]^2$, is a manifestation of this
effect---it represents the
difference between visual and radio refraction.   We assume
that the term's coefficient,
$\xi(\nu)$, is 1 while making submillimeter observations and 0 while making 
optical pointing measurements. 
This is a crude approximation, since $\xi(\nu)$ actually varies 
across the millimeter and submillimeter bands 
in a more complex manner 
which is not well characterized \citep{davis70}.
The errors introduced by this crude approximation
are negligible for AST/RO, however, because the magnitude of the third term is small.
The partial pressure of water vapor is never
higher than 2 mb at South Pole---this 
is the saturation pressure of water vapor \citep{goff46}
at the maximum recorded temperature of  
$- 14 \, \mathrm{C}$ (see Table 1).
The third term cannot therefore change the value of $N$ by more than 6\%,
and this corresponds to a difference between optical and radio
pointing of $9''$ at $20^\circ$ elevation under worst-case conditions.
Equations \ref{eq:refraction} and \ref{eq:index} are approximate at the
level of a few arcseconds \citep{explanatory}.  
Since sources do not change elevation at South Pole,
any residual errors tend to be removed by optical
and radio pointing procedures.

The optical pointing model is determined by measurements of the
positions of stars with a guide telescope mounted 
on AST/RO's elevation structure.  The guide telescope consists
of a 76mm diameter, 750 mm focal length lens
mounted at one end of a 100mm diameter
carbon fiber-reinforced epoxy tube and an Electrim EDC-1000N
CCD camera mounted at the other end on a focussing stage.  
The telescope tube is filled with dry nitrogen gas, sealed, 
warmed with electrical tape, and insulated.  It works well
under all South Pole conditions.
In front of the lens is a deep red filter to facilitate detection
of stars in daylight.  Daylight observing capability is critical
since the telescope is available to engineering and test teams
only during the daylit summer months.  The CCD camera is connected to a
readout and control board in the data acquisition computer.

Following \citet{condon92},
the horizontal pointing error $\Delta Az \, \mathrm{cos}( El )$ can be
expressed as:
\begin{eqnarray}
\Delta Az \, \mathrm{cos}( El ) & = & C_1 + C_2 \, \mathrm{cos}( El ) 
+ C_3 \, \mathrm{sin}( El ) 
+ C_4 \, \mathrm{cos}( El )\, \mathrm{sin}( El ) \nonumber \\
& & \mbox{} + C_5 \, \mathrm{sin}( El )\, \mathrm{sin}( El )
+ C_6 \, \mathrm{cos}( 2 Az )\, \mathrm{cos}( El )
+ C_7 \, \mathrm{sin}( 2 Az )\, \mathrm{cos}( El ) \nonumber \\
& & \mbox{} + C_8 \, \mathrm{cos}( 3 Az )\, \mathrm{cos}( El )
+ C_9 \, \mathrm{sin}( 3 Az )\, \mathrm{cos}( El ) \ \ ,
\label{eq:delaz}
\end{eqnarray}
where $C_1$ is the optical telescope horizontal collimation error,
$C_2$ is the constant azimuth offset (encoder zero point offset),
$C_3$ represents the non-perpendicularity of the elevation and azimuth axes, 
$C_4$ and $C_5$ represent the tilt from vertical of the azimuth axis, and
$C_6$ through $C_9$ represent irregularities in the bearing
race of the azimuth bearing.  The vertical pointing
error $\Delta El$ can be expressed as:
\begin{eqnarray}
\Delta El & = & D_1 + D_2 \, \mathrm{cos}( El ) 
+ D_3 \, \mathrm{sin}( El ) 
+ D_4 \, \mathrm{cos}( Az )\nonumber \\
& & \mbox{} + D_5 \, \mathrm{sin}( Az )
+ D_6 \, \mathrm{cos}( 2 Az )
+ D_7 \, \mathrm{sin}( 2 Az ) \nonumber \\
& & \mbox{} + D_8 \, \mathrm{cos}( 3 Az )
+ D_9 \, \mathrm{sin}( 3 Az ) \ \ ,
\label{eq:delel}
\end{eqnarray}
where $D_1$ is the optical telescope vertical collimation error
plus encoder offset,
$D_2$ and $D_3$ are mechanical sag and elevation encoder decentering,
$D_4$ and $D_5$ 
represent the tilt from vertical of the azimuth axis, and
$D_6$ through $D_9$ represent irregularities in the bearing
race.

A pointing data set consists of position offsets $(\Delta Az_i, \Delta El_i)$ 
as a function of $(Az_i, El_i)$, for several hundred 
observations of stars ($i = 1,\ldots,N$).  
The $\Delta El_i$ values are corrected for optical refraction by
Equations \ref{eq:refraction} and \ref{eq:index}, with $\xi(\nu) = 0$.
The best fit values $C_1,\ldots,C_9$
and $D_1,\ldots,D_9$ in Equations \ref{eq:delaz} and \ref{eq:delel} 
are solved for as
a function of the observed star positions and offsets.  This is an inverse linear
problem solved by standard numerical methods.
The $C_1,\ldots,C_9$
and $D_1,\ldots,D_9$ constants are then inserted into the real-time telescope
drive system control software as corrections to the pointing.  
With this pointing model in place, subsequent observations of stars
show errors of only a few seconds of arc.
This pointing process is repeated yearly.  The high-order terms,
$C_6,\ldots,C_9$ and
$D_6,\ldots,D_9$ have not changed significantly since they were
first measured in 1993.
The low-order terms, 
$C_1,\ldots,C_5$ and
$D_1,\ldots,D_5$ change because of instabilities in the mechanical mounting
of the guide telescope, long-term shifts in the telescope foundations, and
thermal heating of the telescope tower during the Austral summer.
\begin{figure}
\plotone{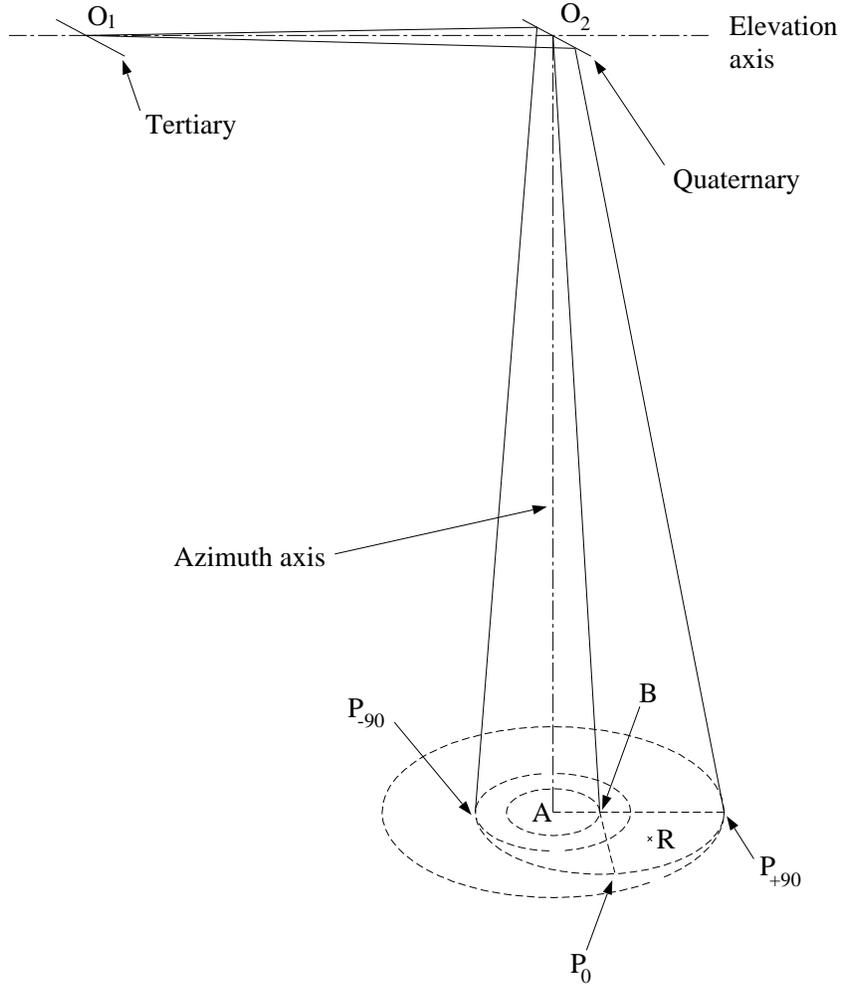}
\bigskip
\caption{{ \bf Geometry of AST/RO optics from the tertiary mirror to
the Coud\'{e} focus \citep[adapted from][]{zhang96}.}
In general, the vectors $\vec{AB}$ and $\vec{BP_{+90}}$ are not parallel,
nor do they 
lie in the same plane as the azimuth and elevation
axes.}
\label{fig:precession}
\end{figure}

\subsection{Radio Pointing}

Implementation of the optical pointing mechanism assures that 
the optical telescope boresight attached to the telescope's 
elevation structure is pointed within specification.  The
radio receivers are, however, located at a Coud\'{e} focus,
which is attached to the telescope mounting tower and
fixed with respect to the earth.  As
shown in Figure \ref{fig:optics}, the tertiary and quaternary
steer the radio beam along the elevation and azimuth axes.
Since the alignment of these mirrors is not perfect, additional
corrections are needed for pointing 
each receiver \citep{zhang96}.  

In Figure \ref{fig:precession}, the elevation axis intersects the
tertiary at $O_1$ and the quaternary at $O_2$.  The intersection
of the mechanical azimuth axis with the Coud\'{e} focal plane is 
point $A$.  A photon 
traveling from $O_1$ to $O_2$ and then reflected from
the quaternary will arrive at the focal plane at a point $B$.
Point $B$ and point $A$ are not
coincident because of misalignment of the quaternary mirror. 
If the tertiary is also misaligned, then the principal ray from
the secondary will not coincide with the elevation axis.
The principal ray reflects from the tertiary and quaternary,
and intersects the focal plane at a point whose
position depends on the elevation angle, $P_{El}$.
Figure \ref{fig:precession} shows that the locus
of points $P_{El}$ describes a circle centered on point $B$.
The three labeled points, $P_{-90}$, $P_{0}$, and $P_{+90}$
show the intersection of the principal ray with the focal
plane at elevation angles of $-90$, $0$, and $+90$.  (Imagine
that the elevation travel includes negative
angles; in actuality
the mechanical stops on AST/RO do not permit negative 
elevation angles.)  Furthermore, when the telescope rotates
around the azimuth axis, the points 
$P_{-90}$, $P_{0}$, $P_{+90}$, and $B$ precess around point $A$
in a fixed pattern.  We therefore see that a star on
the boresight will have an offset from $A$ given by
\begin{equation}
\vec{AP} = (\vec{AB} + \vec{BP_0} {\mathrm e}^{\pm i El}){\mathrm e}^{-i Az},
\end{equation}
where the plus or minus sign in the elevation dependence is determined
by the nature of the tertiary misalignment.

The receivers are not in general aligned with the azimuth axis,
so let point $R$ be the center of the receiver beamwaist at the
Coud\'{e} focus.  The image precession effect as viewed by the
receiver is
\begin{eqnarray}
\vec{RP} & = & \vec{RA} + \vec{AP} = \vec{RA} +
(\vec{AB} + \vec{BP_0} {\mathrm e}^{\pm \, i El}){\mathrm e}^{-i Az}, \nonumber
\\
& = & |\vec{RA}| {\mathrm e}^{i \theta_a} + ( |\vec{AB}| {\mathrm
e}^{i \theta_b} + |\vec{BP_0}|{\mathrm e}^{i[\theta_c\, \pm \, El]} ){\mathrm
e}^{-i Az} \nonumber \\
& = & |\vec{RA}|(\mathrm{cos}\theta_a + i \, \mathrm{sin}\theta_a) +
|\vec{AB}|\mathrm{cos}(\theta_b - Az) + i \, |\vec{AB}|\mathrm{sin}(\theta_b - Az)
\nonumber \\
&  & + |\vec{BP_0}|\mathrm{cos}(\theta_c \pm El - Az) + i \, |\vec{BP_0}|\mathrm{sin}(\theta_c \pm El - Az) , 
\end{eqnarray}
where $\theta_a$, $\theta_b$, and $\theta_c$ are the phase angles of
vectors $\vec{RA}$, $\vec{AB}$, and $\vec{BP_0}$, respectively.
The Cartesian components, $x$, $y$, of the focal plane image viewed by
the receiver changes with the azimuth and elevation of the source as
\begin{equation}
x_{\vec{RP}} = |\vec{RA}|\mathrm{cos} \theta_a +
|\vec{AB}|\mathrm{cos}(\theta_b + Az) +
|\vec{BP_0}|\mathrm{cos}(\theta_c \, \pm \, El - Az) ,
\label{eq:xrp}
\end{equation}
and
\begin{equation}
y_{\vec{RP}} = |\vec{RA}|\mathrm{sin} \theta_a +
|\vec{AB}|\mathrm{sin}(\theta_b + Az) +
|\vec{BP_0}|\mathrm{sin}(\theta_c \, \pm \, El - Az) .
\label{eq:yrp}
\end{equation}
These coordinates represent the point within the image plane
of the Coud\'{e} focus which is observed by the receiver.

The radio beam pointing corrections modify the pointing of
the telescope so that the nominal boresight falls on a particular
receiver at all azimuth and elevation.  These corrections are
an inversion of the effect described by Equations \ref{eq:xrp}
and \ref{eq:yrp}:
\begin{equation}
\Delta Az \, \mathrm{cos}(El) = A_1 \mathrm{cos}(\theta_1 \mp El + Az) +
A_2 \mathrm{cos}(\theta_2 \mp El) + A_3 \mathrm{cos}(\theta_3) ,
\label{eq:rdelaz1}
\end{equation}
and 
\begin{equation}
\Delta El = A_1 \mathrm{sin}(\theta_1 \mp El + Az) + A_2
\mathrm{sin}(\theta_2 \mp El ) + A_3 \mathrm{sin}(\theta_3) ,
\label{eq:rdelel1}
\end{equation}
where $A_1 \mathrm{e}^{i \theta_1}$ corrects the error due to the
misalignment of the receiver, $A_2 \mathrm{e}^{i \theta_2}$ corrects
the misalignment of the quaternary, and $A_3 \mathrm{e}^{i \theta_3}$
corrects for the combined misalignment of the primary, secondary, and
tertiary plus any collimation offsets between the optical guide
telescope and the radio beam.  In this formulation, there
are six constants and the unknown sign of the elevation dependence
to be determined by fits to radio pointing data.  As a practical
matter, it is easy with AST/RO to obtain accurate radio data
for a few bright sources at all azimuths, but known sources suitable 
for radio pointing are only found at a few elevations.  Equations
\ref{eq:rdelaz1} and \ref{eq:rdelel1} are therefore modified to
absorb the elevation dependence into the fitting constants:
\begin{equation}
\Delta Az \, \mathrm{cos}(El) = B_1(El) \, \mathrm{cos}[\theta_1(El) + Az] +
B_2(El),
\label{eq:rdelaz2}
\end{equation}
and
\begin{equation}
\Delta El = B_3(El) \, \mathrm{sin}[\theta_2(El) + Az] + B_4(El) .
\label{eq:rdelel2}
\end{equation}
For each elevation, a series of radio maps is used to determine
the six parameters $B_1$, $B_2$, $B_3$, $B_4$, $\theta_1$, and
$\theta_2$.  If the mount errors have been correctly compensated by
Equations \ref{eq:delaz} and \ref{eq:delel}, then 
$B_1 = B_3$ and $\theta_1 = \theta_2$.
When these parameters are measured at several
elevations, their value at all elevations is estimated by
linear or quadratic interpolation in $El$.
The magnitude of the pointing corrections given by Equations \ref{eq:rdelaz2}
and \ref{eq:rdelel2} are typically a minute of arc.

\subsection{Chopper Offsets}

\begin{figure}
\epsscale{0.8}
\plotone{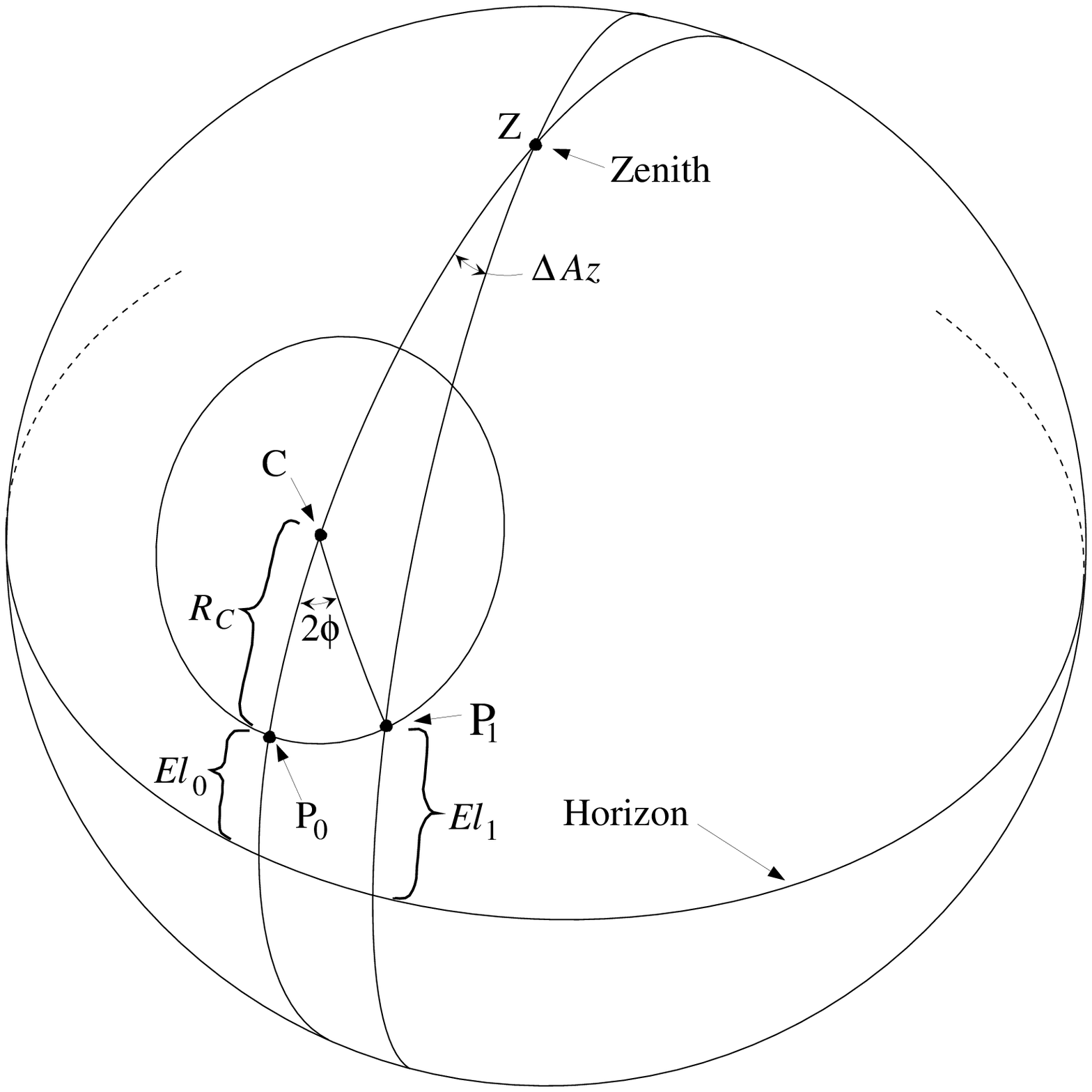}
\bigskip
\caption{{ \bf Geometry of
AST/RO chopper motion on the celestial sphere.}  
Point Z indicates the zenith of the celestial sphere, and
point $\mathrm{P_0}$ is the direction of the telescope beam
when the chopper is centered.  When the chopper mirror
rotates through an angle $\phi$, the beam center moves to
point $\mathrm{P_1}$ along a small circle centered at C.
The radius of this circle is the spherical angle $R_C$,
whose value depends on the geometry of the optics.
}
\label{fig:choppergeom}
\end{figure}

\begin{figure}
\epsscale{0.8}
\plotone{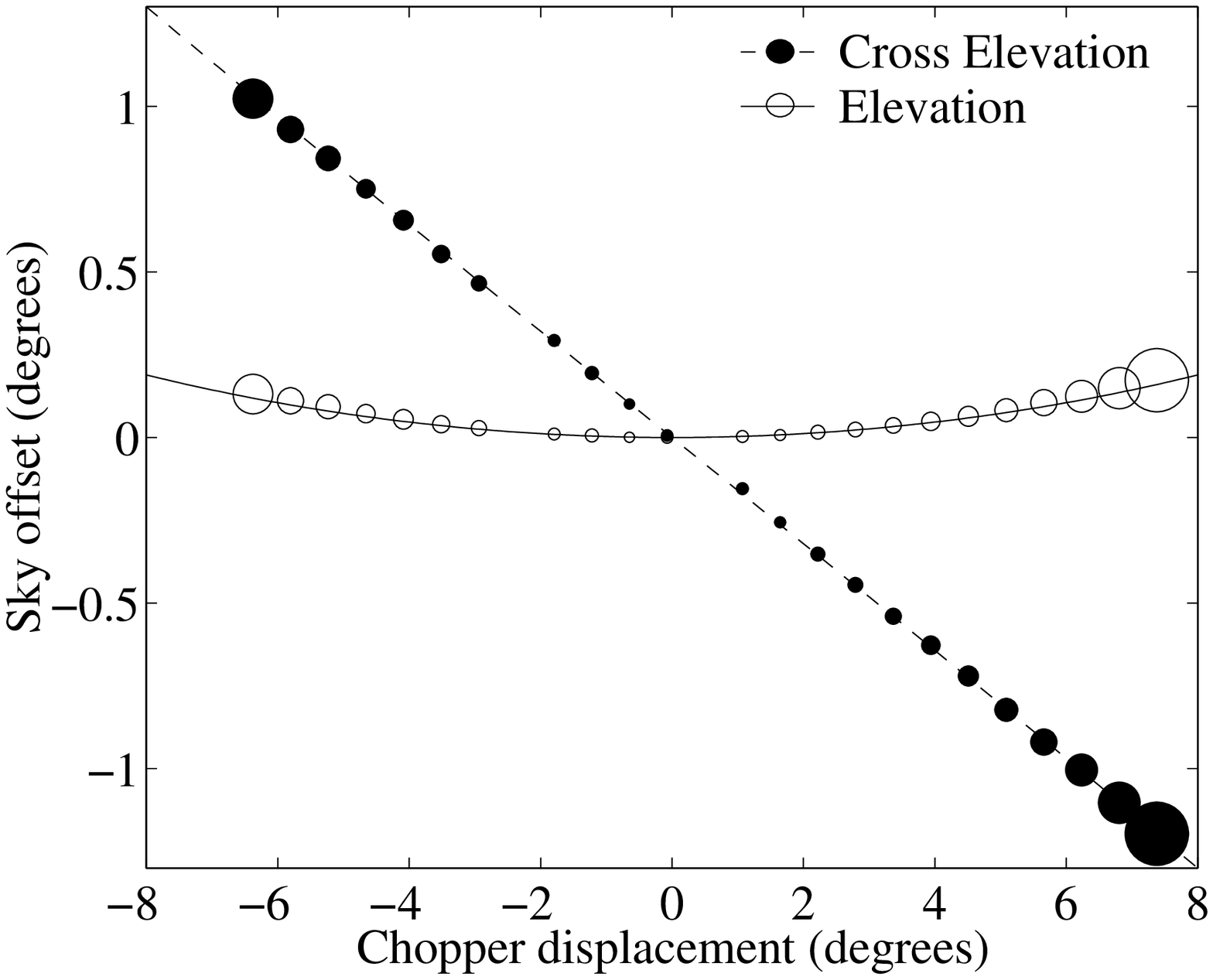}
\bigskip
\caption{{ \bf Measurements of the 
AST/RO chopper.}  
Beam position is measured as a function of chopper
displacement.  Shown are elevation and cross-elevation
pointing offsets as a function of chopper motion.  The
ellipses represent beam size and position data measured
during scans of the Sun at $\lambda 600 \mu \mathrm{m}$; 
each ellipse is similar
in shape to the beamshape at that chopper displacement,
and the size of each ellipse is given by the scale on
the ordinate axis.
The curves are calculations of the beam centroid based on a
computer model of the optics.   
}
\label{fig:chopper}
\end{figure}

As the chopper rotates,
the beam position moves in both azimuth and elevation.
Figure \ref{fig:choppergeom} 
shows the geometry of the beam position on the celestial
sphere.
When the chopper mirror is in its nominal, centered position,
the radio beam points toward ${\mathrm{P_0}}$.
When the chopper mirror rotates through an angle $\phi$, the
beam rotates through an angle $2 \phi$ and moves on the sky
to point ${\mathrm{P_1}}$.
The resulting change in elevation and azimuth
can be calculated by considering the spherical triangle
${\mathrm{ZCP_1}}$, where Z is the zenith and C is the
center of a small circle on the sky whose radius is the
spherical angle $R_C$.
Solving for the arc ${\mathrm{ZP_1}}$,
the spherical law of cosines gives:
\begin{equation}
\mathrm{sin} (El_1) = \mathrm{sin} (R_C + El_0) \, \mathrm{cos} (R_C)  - 
\mathrm{cos} (R_C + El_0) \, \mathrm{sin} (R_C) \, \mathrm{cos} (2\phi) ~~~,  
\end{equation}
where $El_0$ and $El_1$ are the elevations of 
${\mathrm{P_0}}$ and ${\mathrm{P_1}}$.
The spherical law of sines gives:
\begin{equation}
\mathrm{sin} (\Delta Az) = - {{\mathrm{sin} (2 \phi) \, 
\mathrm{sin} (R_C) } \over {\mathrm{cos} (El_1) }} ~~~,  
\end{equation}
where $\Delta Az$ is the change in azimuth.
These  equations hold even if $El_0 + R_C > 90^o$, moving point C
over-the-top to the other side of the zenith.
In the AST/RO system,
the value of $\phi$ is given by the encoder on the chopper motor.
The value $R_C = 4\,.\!\!^o 596$ has been determined by measurements
of the Moon and Sun. 
Figure \ref{fig:chopper} shows the beamsize and displacement
as a function of chopper angle.

\subsection{Data Calibration}

The real brightness temperature of an astronomical source 
depends only on the intrinsic radiation properties of the 
source, whereas the measured antenna temperature is influenced by the 
efficiency of telescope-to-source radiation coupling and the 
conditions in the Earth's atmosphere through which the radiation must 
propagate.  
Calibration converts from spectrometer output counts in the
data acquisition computer
to the source effective radiation temperature, correcting for 
atmospheric losses.
The intermediate frequency (IF) output of a mixer in one
of the AST/RO receivers is 
comprised of down-converted signals from both the upper and lower 
sidebands.  The amplified IF signal is detected by an acousto-optical 
spectrometer (AOS).  The result is a spectrum of data with values proportional
to the incident power spectrum.  Discussed in this section is the 
conversion of these data values to the {\em radiation temperature} scale,
$T_R^{*}$.

The {\em chopper wheel} method of calibration was originally
described by \citet{penzias73} and elaborated upon by
\citet{ulich76} and \citet{kutner81}.
The theory of absolute calibration of millimeter-wavelength 
data has not changed in its
fundamentals since those papers were written, but the
introduction of automated calibration loads at temperatures other
than ambient has led to improvements in technique and a 
conceptual shift in the understanding of calibration parameters.
\begin{itemize}
\item{The ambient chopper-wheel calibration method determined
the relative gain of each spectrometer channel at the time
of the chopper-wheel measurement, and later applied that
gain to the switched astronomical data, assuming that it
had not changed.  Data
quality is improved by measuring the relative gain
during the acquisition of the astronomical data 
using the ``{$(S - R)/R\,$}'' technique, where $S$ refers to the
``on-source'' (or signal) measurement and $R$ refers to an ``off-source''
(or reference) measurement.  Calibration of such
a spectrum requires, however, determination of
the {\em atmosphere-corrected system 
temperature}, $T_{\mathrm{sys}}^{*}$.
}
\item{The {\em receiver temperature}, $T_{\mathrm{rx}}$,
and {\em sky temperature}, $T_{\mathrm{sky}}$, are distinct
quantities measuring different sources of noise in the system.
The former is a property of the instrument and
under the control of the observer, while the latter
is a property of the ever-changing sky.  
Using the ambient chopper-wheel method as originally
formulated \citep{penzias73}, 
these two noise power levels cannot be determined independently,
because there is only a single
calibration load of known temperature; this means, for example, 
that it is not possible to determine whether 
a loss of observing sensitivity is caused by a
problem with the receiver or an increase in atmospheric
opacity.  With multiple calibration loads, as in the AST/RO
system, both of these quantities can be determined and the
observing program modified as appropriate.}
\end{itemize}

\paragraph{Relation between $\ta$\ and $\Tb$.}

The {\it antenna temperature} for an astronomical source is defined as its
brightness temperature in the Rayleigh-Jeans limit.  If a radio telescope
observes a source of specific intensity $I_\nu$, the source
brightness temperature $\Tb$\ at frequency $\nu$ is given by the Planck function,
\begin{equation}
I_\nu ~\stackrel{\Tb}{\equiv}~ {2h\nu^3\over{c^2}}{1\over{\exp(h\nu/k\Tb) - 1}}~~~.\label{tbdef}
\end{equation}
The blackbody emitter at temperature $\Tb$ described by this equation
is often fictitious---the usual circumstance for submillimeter
observations of the interstellar medium is that 
the observed flux, $I_\nu$, results from
thermal processes in emitters at temperatures higher than $\Tb$ that
do not fill the telescope beam or are not optically thick.
The antenna temperature is given by the Rayleigh-Jeans limit to
Equation \ref{tbdef},
\begin{equation}
\label{tadef}
I_\nu ~\stackrel{\ta}{\equiv}~ {2k\nu^2\ta\over{c^2}}~~~.
\end{equation}
The {\it effective radiation temperature} is
\begin{equation}
\label{jnudef}
J_\nu(T) ~\equiv~ {h\nu/k\over{\exp(h\nu/\ k T) - 1}}~~~.
\end{equation}
In the Rayleigh--Jeans limit, ($h\nu/k~<<~T$), $J_\nu(T) = T \, $;
this approximation applies to essentially all temperatures
within the AST/RO system, since $h \cdot 1 \, \mathrm{THz} /k = 4.8 \, \mathrm{K}$,
which is considerably lower than any physical temperature found in the
AST/RO calibration system, telescope, or the atmosphere.
It must be remembered, however, that the small flux levels of
observed astronomical sources are expressed in the $\ta$
scale, which is simply defined to be proportional to power: $\ta =
(c^2/2k\nu^2) \, I_\nu$.   This is a linear extrapolation to low power
levels from the higher power levels of the calibration loads,
where the calibration loads do satisfy the Rayleigh--Jeans approximation
at these frequencies.  
Setting Equations \ref{tbdef} and \ref{tadef} equal, we see that
antenna temperature is the effective radiation temperature of the
source brightness temperature,
\begin{equation}
\label{tajnu}
\ta  =  J_\nu(\Tb)~~~,
\end{equation}
but in general $\ta$ is small 
so that $\Tb \mathrel{\raise0.35ex\hbox{$\scriptstyle
>$}\kern-0.6em\lower0.40ex\hbox{{$\scriptstyle \sim$}}} \ta$.

\paragraph{Calibration scans.}

The AST/RO 
calibration system allows the receivers to ``chop'' between three blackbody
loads and the sky \citep[cf. Figure \ref{fig:optics} and ][]{stark97a}.  There is a load
at ambient receiver room temperature (the ``warm" load), and two loads cooled by a closed
cycle refrigerator to 40 K (the ``cold'' load) and 90 K  (the ``cool'' load).  
The surfaces of the refrigerated loads are warmed 
to temperatures higher than their average physical temperature 
by infrared radiation entering the dewar windows, 
so that their effective radiation temperatures are
about 100 K and 140 K, respectively. 
The effective radiation temperatures of the loads 
are measured about once a month 
by manually comparing the
receiver response to each load with that of an absorber soaked in liquid 
nitrogen and to the warm load.  
The physical temperatures of all the loads are monitored by the computer,
and we assume that if the physical temperature has not changed,
the effective radiation temperature has not changed either.

During a calibration scan (performed via the OBS command {\tt ca}), the
data acquisition computer records data for
each AOS channel corresponding to the receiver response to (1) no
incoming IF signal ({\it zero} measurement), $D_{{\mathrm{zero}j}}$ 
(this is essentially the dark current of 
the CCD output stage of the AOS),
(2) the ``cold''
calibration load, $D_{\mathrm{cold}j}$, 
and (3) the ``warm'' calibration load,  
$D_{\mathrm{warm}j}$. 
The ``cool'' load can optionally be used in place of
either the ``cold'' or ``warm'' loads.  
It is best if the power levels
of the receiver with the loads in place during
calibration is similar
to the power level while observing the sky.
The AOS operates as a linear
power detector,
i.e. the spectrum of measured
data values $D_j$ is proportional to the incident power spectrum.  For the 
$j$th AOS channel, we measure the {\it gain}:
\begin{equation}
\Gamma_j  =  { {D_{{\mathrm{warm}j}} - D_{{\mathrm{cold}}j} }\over 
{J_{\nu}({ T_{\mathrm{warm}}}) - J_{\nu}({ T_{\mathrm{cold}}})} }~~~.
\label{gains}
\end{equation}
The gains spectrum gives us the proportionality between 
the antenna temperature scale $(\ta)$ and
the arbitrary
intensity counts ($D$) read out from the AOS 
by the computer.
The {\it calibration spectrum} 
measures the noise of the receiver, IF system, and AOS:
\begin{equation}
T_{\mathrm{rx}j}  =  { {D_{{\mathrm{cold}j}} - D_{{\mathrm{zero}}j}}\over \Gamma_j} - 
J_{\nu}({ T_{\mathrm{cold}}})~~~.
\label{cal}
\end{equation}
The {\it average receiver noise temperature}, 
${ \langle T_{\mathrm{rx}}\rangle}$, is the average
of the calibration spectrum,
\begin{equation}
{\langle  T_{\mathrm{rx}}\rangle} ~ \equiv ~ 
{{1}\over{N_c}} \, \sum_{j=1}^{N_c} T_{{\mathrm{rx}}j}~.
\label{rxave}
\end{equation}
where the average is performed over a selectable subset 
of spectrometer channels.

\begin{figure}
\plotone{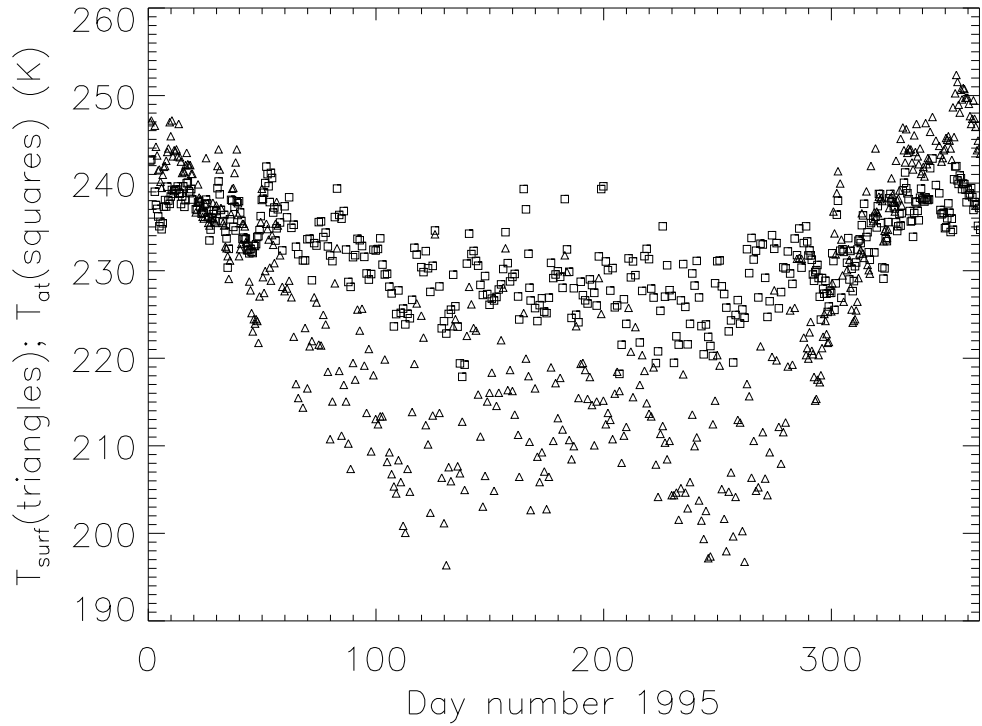}
\bigskip
\caption{{ \bf $T_{\mathrm{at}}$ and $T_{\mathrm{surf}}$ vs. time at South Pole}
\citep{ingalls99}.
Note the temperature inversion which occurs during the polar winter
($T_{\mathrm{at}}$ $>$ $T_{\mathrm{surf}}$).
}
\label{fig:tscatter}
\end{figure}

\begin{figure}
\plotone{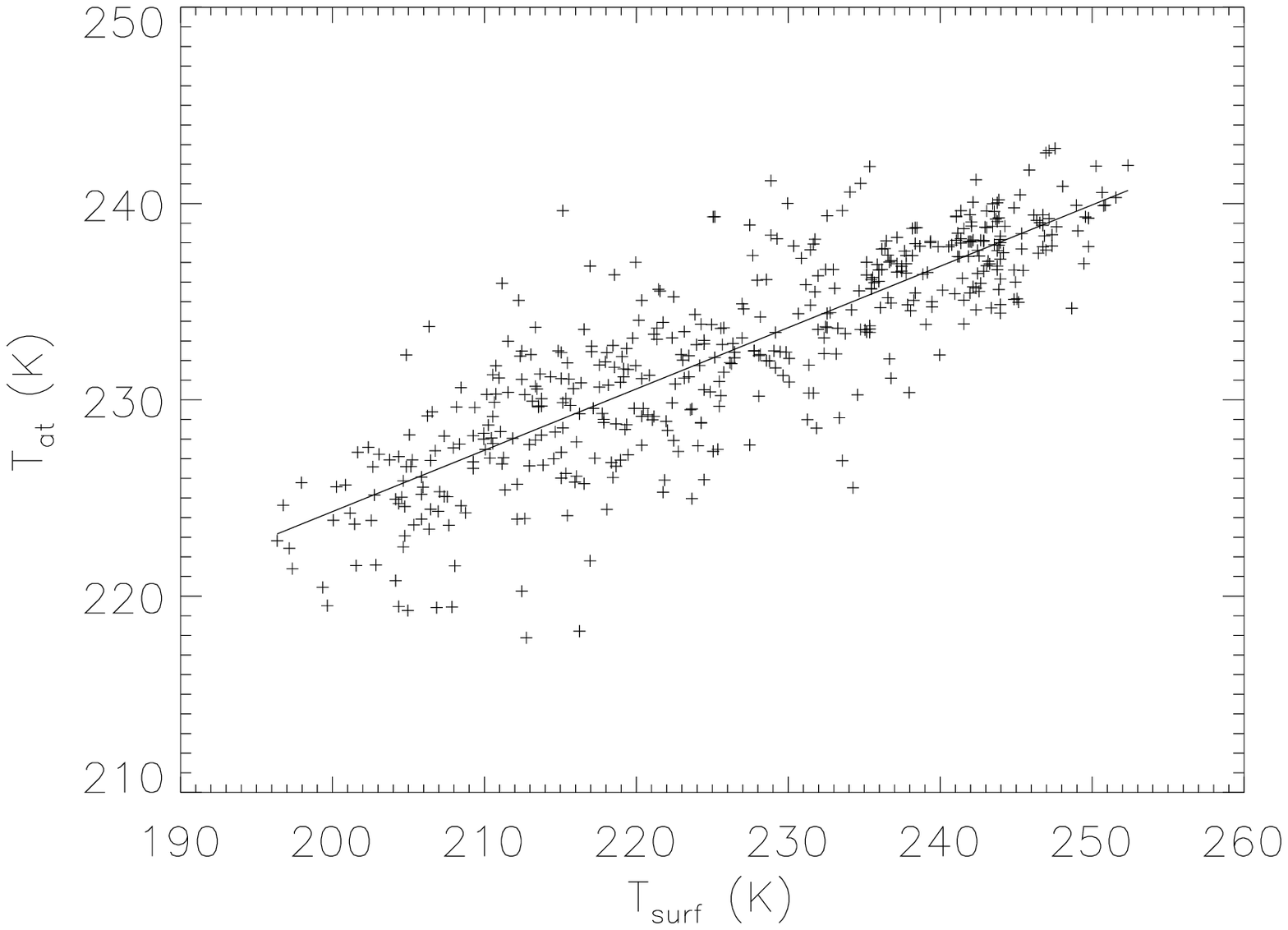}
\bigskip
\caption{{ \bf $T_{\mathrm{at}}$ vs. $T_{\mathrm{surf}}$ at South Pole}
\citep{ingalls99}.
The solid line shows the fit $T_{\mathrm{at}} = 0.31\, T_{\mathrm{surf}} + 162\, \mathrm{K}$.
}
\label{fig:tsurf}
\end{figure}

\paragraph{The single-slab model of the atmosphere.}
In the {\it single-slab model}, the atmosphere is assumed to be
a plane-parallel medium
with mean temperature $\tat$ and double-sideband zenith 
opacity $\tau_{\mathrm{dsb}}$.
The sky brightness temperature can be expressed as \citep{chamberlin97}
\begin{equation}
\tsky  =  T_{\mathrm{spill}} + \eta_l J_\nu(\tat)\left[ 1 - 
\exp(-\tau_{\mathrm{dsb}} A)\right]~~~,
\label{skydip}
\end{equation}
where we have introduced the {\it spillover temperature}, $T_{\mathrm{spill}}$,
the {\it telescope loss efficiency}, $\eta_l$, 
and $A \equiv \mathrm{csc}(El)$ is the airmass at the elevation of the
telescope.
The effective radiation temperature of the cosmic microwave 
background radiation is ignored because it is negligible at 
submillimeter wavelengths.
The telescope loss efficiency,
$\eta_l$, is the fraction of receiver response that originates
from the sky. 
The spillover temperature, $T_{\mathrm{spill}}$, is the effective radiation 
temperature of
all sources of radiation incident on the receiver which are not from the
direction of the sky.  These sources primarily include the thermal emission 
from the optics and spillover from the Earth's surface.  
$T_{\mathrm{spill}}$ can be expressed as
\begin{equation}
\label{Tspill}
T_{\mathrm{spill}} = (1 - \eta_l) J_\nu (T_\mathrm{sbr}) ~,
\end{equation}
where $T_\mathrm{sbr}$ is by definition the average temperature of the emitters
giving rise to $T_{\mathrm{spill}}$ \citep{ulich76}.
Since there is a gradient in temperature between the air
outside of the telescope and the warmed environment
of the receiver room,
$T_\mathrm{sbr}$ will fall somewhere between the 
outside surface temperature, 
$T_\mathrm{surf} \sim 230 \, \mathrm{K}$, and the
temperature of the receiver room, 
$T_\mathrm{warm} \approx 293 \, \mathrm{K}$, depending on
the quality of the receiver feed and its alignment.

\paragraph{Sky brightness measurement.}
The atmospheric attenuation of the signal is estimated by measuring
the sky brightness temperature
$\tsky$ .  The {\it sky spectrum} is made by measuring the sky and a cold
load using the {\tt sk} command in OBS.  These are combined to yield:
\begin{equation}
T_{{\rm sky}j} =  { {D_{{\mathrm{sky}j}} - D_{{\mathrm{cold}j}}}
\over \Gamma_j} + J_{\nu}({ T_{\mathrm{cold}}})~~~,
\label{sky}
\end{equation}
and averaging over channels gives
\begin{equation}
\langle \tsky \rangle ~ \equiv ~ {{1}\over{N_c}} \, \sum_{j=1}^{N_c} T_{{\rm sky}j}~.
\label{skyave}
\end{equation}

A {\it skydip} measures $\langle \tsky \rangle$ for a range of 
different elevations.  These data can be fit to the single-slab 
model (Equation \ref{skydip}) for the 
terms $T_{\mathrm{spill}}$ and $\tau_{\mathrm{dsb}}$, and for the product 
$\eta_l J_\nu (\tat)$ (Chamberlin \etal\ 1997), which is
used to determine $\eta_l\,$.  

Instead
of assuming that the mean atmospheric temperature, $\tat$, is equal to the ambient
temperature, $T_\mathrm{surf}$, which is standard chopper-wheel calibration procedure,
we use actual {\it in situ} atmospheric measurements to deduce the
mean atmospheric temperature.
The South Pole meteorological station launches an upper air balloon 
once or twice each day to measure
air temperature and relative humidity as a function of 
height.
These data can be used to derive $\tat$, the mean 
atmospheric temperature in the column of air above the telescope.  For
each balloon flight in 1995 and 1996, we have approximated
$\tat$ at $\nu = 492 \, \mathrm{GHz}$ 
by averaging the physical temperature of
the atmosphere, $T$, over altitude, weighted by the 
opacity of seven nearby water lines:
\begin{equation}
\tat \simeq {{\sum_{k=1}^{7}  \int_0^{z_{\mathrm{max}}}T \, n_{\rm H_2O} \,
S_k(T) F_k(\nu_k,\nu) \mathrm{d}z\over
{\sum_{k=1}^{7} \int_0^{z_{\mathrm{max}}}n_{\rm H_2O} \,
S_k(T) F_k(\nu_k,\nu) \mathrm{d}z}}}, \label{tat}
\end{equation}
where $n_{\rm H_2O}$ is the density of water vapor at height $z$,
$\nu_k$ is the frequency of water line $k$,
$S_k (T)$ is the line intensity of water line $k$, and $F_k(\nu_k,\nu)$
is a Lorentzian line-broadening function.
During the South Pole winter, a temperature inversion occurs where
the mean atmospheric temperature, $\tat$, is higher than the surface
temperature, $\tsurf$ .  Figure \ref{fig:tscatter} is a plot 
of $\tat$ and $\tsurf$
vs. time for the year 1995.  Despite the presence
of this inversion layer for only part of the year, $\tat$ is 
functionally related to $\tsurf$, with little scatter.
Figure \ref{fig:tsurf} shows $\tat$ vs. $\tsurf$ for 1995, with a straight 
line fit to the data.  
The fit to the 1996 data is the same.
Adopting the relationship
\begin{equation}
\tat  \approx  0.3 \, \tsurf + 162 \,  {\rm K}
\label{surfat}
\end{equation}
represents ninety percent of the data within $\pm \,4\,$K.
Since $\tat \approx 230 \,$K, this is a spread of only $\pm 2\%$.
Given a value of the surface temperature when a skydip measurement
was made, this equation gives an estimate for $\tat$. 
The product $\eta_l J_\nu (\tat)$, fit to skydip data, then gives an estimate 
of the telescope loss efficiency $\eta_l$.  This procedure is most
accurate during the best weather, when the fit to
Equation \ref{skydip} is best.  We assume that for a given 
configuration of the optics
and receivers, $\eta_l$ is constant and independent of the weather.
Values of $\eta_l$ on AST/RO have ranged from 0.65 to 0.92 for various
alignments of the various receivers.

\paragraph{Single and Double Sideband Atmospheric Opacity.}
For receiver systems with no method of sideband rejection,
a given measurement is necessarily a {\it double--sideband} measurement.
In other words, the measured response of an AOS channel is a
combination of 
{\sl two different frequencies},
the {\it signal} sideband and the {\it image} sideband, possibly with
different receiver gain in the two sidebands.  
The gains in the two sidebands are normalized so that 
$g_{\mathrm{s}} + g_{\mathrm{i}} = 1$.  All of the current AST/RO receivers are double sideband,
with $g_{\mathrm{s}} \cong g_{\mathrm{i}} \cong {{1}\over{2}}$.

A double--sideband measurement of the sky brightness temperature
is weighted by the gain, as well as the atmospheric opacity, in each sideband:
\begin{equation}
\tsky =  g_{\mathrm{s}}T_{\mathrm{spill},{\mathrm{s}}} 
   + g_{\mathrm{i}}T_{\mathrm{spill},{\mathrm{i}}} + 
\eta_l J_\nu(\tat) \left[ g_{\mathrm{s}}(1 - 
{\rm e}^{-\tau_{\mathrm{s}} A}) + g_{\mathrm{i}}(1 - {\rm e}^{-\tau_{\mathrm{i}} A}) \right]~~~,
\label{dsbskydip}
\end{equation}
where $\tau_{\mathrm{s}}$ and $\tau_{\mathrm{i}}$ are the zenith opacities in the signal and
image sidebands, respectively, and $T_{\mathrm{spill},{\mathrm{s}}}$ 
and $T_{\mathrm{spill},{\mathrm{i}}}$ are the
spillover radiation temperatures at the frequencies $\nu_s$ and $\nu_i$,
respectively.  Comparing Equations
(\ref{dsbskydip}) and (\ref{skydip}), we see that we can use the
skydip fitting method if 
\begin{equation}
T_{\mathrm{spill}} ~ \equiv ~ g_{\mathrm{s}}T_{\mathrm{spill},{\mathrm{s}}} + 
g_{\mathrm{i}}T_{\mathrm{spill},{\mathrm{i}}}
\end{equation}
and
\begin{equation}
{\rm e}^{-\tau_{\rm {\mathrm{dsb}}} A} ~ \equiv ~ g_{\mathrm{s}}{\rm e}^{-\tau_{\mathrm{s}} A} + g_{\mathrm{i}}{\rm e}^{-\tau_{\mathrm{i}} A}~~~.
\label{dsbtau}
\end{equation}

\paragraph{Spectral Line Measurements.}

The spectral line appears in only one of the sidebands.  
The source antenna temperature is a sum of
contributions along the line of sight from both the source and the sky.   
Following \citet{ulich76}:
\begin{eqnarray}
{T_{{\mathrm{source}j}}} ~& = &~ g_{\mathrm{s}} [ T_{\mathrm{spill},{\mathrm{s}}}~ +~ \eta_l J_\nu (\tat)
(1 ~-~ {\rm e}^{-\tau_{\mathrm{s}} A}) +~ \eta_l T_{Aj}^{*} ~{\rm e}^{-\tau_{\mathrm{s}} A} ] 
\label{tsource} \\
                        & ~ & +~ g_{\mathrm{i}}\left [ T_{\mathrm{spill},{\mathrm{i}}}~ +~ 
\eta_l J_\nu(\tat) (1 ~-~ {\rm e}^{-\tau_{\mathrm{i}} A}) \right ]~~~,\nonumber
\end{eqnarray}
\noindent where $T_{Aj}^{*} \,$ is the corrected antenna temperature of
spectrometer channel $j$.
Again, the cosmic microwave background radiation is ignored.

Observing always involves a switching scheme, where the frequency or
position is changed and the signal on blank sky is subtracted from
the signal on the source plus blank sky.
In other words, subtracting Equation 
\ref{dsbskydip} from Equation \ref{tsource} gives:
\begin{eqnarray}
{T_{Aj}} & ~\equiv~ & {T_{{\mathrm{source}j}}}~ - ~ T_{\mathrm{sky}j} 
\label{tadef2} \\
                             &  =     & g_{\mathrm{s}}\eta_l T_{Aj}^{*} ~{\rm e}^{-\tau_{\mathrm{s}} A}, 
\label{dtsource}
\end{eqnarray}
or
\begin{equation}
T_{Aj}^{*} = {{{\mathrm{e}}^{\tau_{\mathrm{s}} A}}\over{g_{\mathrm{s}}\eta_l}}\, 
{T_{Aj}} ~\stackrel{M}{\equiv}~ M \, {T_{Aj}} ~~~,
\label{multiplier}
\end{equation}
where $M$ is the multiplier relating corrected and uncorrected antenna temperature.
Skydips and sky temperature measurements yield 
$\tau_{\mathrm{dsb}}$, not $\tau_{\mathrm{s}}$.
Substitution of Equation \ref{dsbtau} into Equation \ref{multiplier} shows that
\begin{equation}
M \equiv {{{\mathrm e}^{\tau_{\mathrm{s}} A}}\over{g_{\mathrm{s}}\eta_l}} = 
{{{\mathrm{e}}^{\tau_{\mathrm{dsb}} A}}\over{\eta_l}}  \left
 [ {{1}\over{1 - g_{\mathrm{i}} \, 
{\mathrm{exp}} \left ( A [\tau_{\mathrm{dsb}} - \tau_{\mathrm{i}}] \right ) }} \right ]  ~~~.
\label{multiplier2}
\end{equation}
The term $ [\tau_{\mathrm{dsb}} - \tau_{\mathrm{i}}]$
is approximately constant in time and independent of PWV, since 
at most frequencies it is the dry-air opacity which varies rapidly with
frequency, whereas the water-vapor pseudocontinuum opacity varies slowly with
frequency (cf. Figure \ref{fig:pardo}).
This term can therefore be treated as a constant correction term for
a given tuning.
One frequency and tuning at which
this correction is particularly important is 492.1607 GHz \ci,
with the receiver tuned upper sideband so that the image sideband
is near the oxygen line at 487.25 GHz.  In this case, our 1.5 GHz  IF
frequency gives:
$\tau_{\mathrm{dsb}} - \tau_{\mathrm{i}} \cong - 0.27~~~$.

In the AST/RO calibration system, the sky spectrum measurements
are made several times each hour and are used to estimate the
atmospheric opacity.  Equations \ref{skydip} and \ref{Tspill}
can be rearranged to give:
\begin{equation}
{{\mathrm{e}^{\tau_{\mathrm{dsb}} A}}\over{\eta_l}} = 
{{J_\nu(T_\mathrm{at}})\over
{\eta_l J_\nu(T_\mathrm{at}) + 
(1-\eta_l) J_\nu(T_\mathrm{sbr}) - 
\langle T_\mathrm{sky} \rangle}}  ~.
\label{skytau}
\end{equation}
Since $T_\mathrm{at} \approx
T_\mathrm{sbr} \approx T_\mathrm{surf}$, the denominator
of this expression is essentially $T_\mathrm{surf} -
\langle T_\mathrm{sky} \rangle$, with corrections.  
When the opacity is high and $M$ is large, as often
occurs in submillimeter observations, these
corrections are important.
Note that the entire expression depends only weakly on $\eta_l \,$.
We therefore see that $M$ can be expressed in terms of known
quantities:
\begin{equation}
M = 
{{J_\nu(T_\mathrm{at}})\over
{\eta_l J_\nu(T_\mathrm{at}) + 
(1-\eta_l) J_\nu(T_\mathrm{sbr}) - 
\langle T_\mathrm{sky} \rangle}} \,
\left [ {{1}\over{1 - g_{\mathrm{i}} \,
{\mathrm{exp}} \left ( A [\tau_{\mathrm{dsb}} - \tau_{\mathrm{i}}] \right ) }} \right ]  ~~~.
\label{multiplier3}
\end{equation}
For each spectrum, we define the {\em atmosphere-corrected system temperature}:
\begin{equation}
T_\mathrm{sys}^{*} \equiv M
(\langle T_\mathrm{sky} \rangle + \langle T_\mathrm{rx} \rangle ) ~~~,
\label{Tsys}
\end{equation}
where $\langle T_\mathrm{rx} \rangle$ has been determined
by the most recent calibration 
and $\langle T_\mathrm{sky} \rangle$ has
been determined by the most recent
sky brightness measurement at or near the elevation of the
source.
The atmosphere-corrected system temperature is the noise
power in a hypothetical 
telescope system above the earth's atmosphere which
has  sensitivity equivalent to ours.
$T_\mathrm{sys}^{*}$ values for AST/RO vary from $\sim 200 \, \mathrm{K}$
at 230 GHz to $\sim 30,\! 000 \, \mathrm{K}$ at 810 GHz, depending on
receiver tuning and weather.

%
%

\paragraph{Spectral Line Data Acquisition.}
When observing the source, the average output of the $j^{\mathrm{th}}$
channel of the spectrometer is
\begin{equation}
D_{\mathrm{source}j} = \Gamma_j (T_{\mathrm{source}j} + T_{\mathrm{rx}j})
+ D_{\mathrm{zero}j} ~~~,
\label{Dsource}
\end{equation}
and when observing the reference, it is
\begin{equation}
D_{\mathrm{ref}j} = D_{\mathrm{sky}j} = 
\Gamma_j (T_{\mathrm{sky}j} + T_{\mathrm{rx}j})
+ D_{\mathrm{zero}j} ~~~.
\label{Dsky}
\end{equation}
These quantities can be combined using 
Equation \ref{tadef2} to yield a measurement of the
antenna temperature of the source:
\begin{equation}
T_{Aj} = {{D_{\mathrm{source}j} - D_{\mathrm{ref}j}}\over
{D_{\mathrm{ref}j} - {D_{\mathrm{zero}j}}} }
(T_{\mathrm{sky}j} + T_{\mathrm{rx}j}) ~~~,
\label{Taj}
\end{equation}
where 
$T_{\mathrm{sky}j}$
has been determined by the previous sky measurement ({\tt sk}, Equation
\ref{sky}), and
$T_{\mathrm{rx}j}$ and 
$D_{\mathrm{zero}j}$
have been determined by the previous calibration measurement ({\tt ca}, Equation
\ref{cal}).

A better quality spectrum can be obtained by replacing
$T_{\mathrm{sky}j}$ and $T_{\mathrm{rx}j}$ in Equation \ref{Taj} 
by their average
quantities
$\langle T_\mathrm{sky}\rangle$ and 
$\langle T_\mathrm{rx}\rangle$,
from Equations \ref{rxave} and \ref{skyave}:  
\begin{equation}
T_{Aj} \cong {{D_{\mathrm{source}j} - D_{\mathrm{ref}j}}\over
{D_{\mathrm{ref}j} - {D_{\mathrm{zero}j}}} }
(\langle T_\mathrm{sky} \rangle + \langle T_\mathrm{rx} \rangle ) ~~~.
\label{Taj2}
\end{equation}
This is because the actual noise power
in the sky and from the receiver vary slowly across the bandpass
of the spectrometer, whereas the calibration and sky spectra 
$T_{\mathrm{rx}j}$ and 
$T_{\mathrm{sky}j}$
show
spurious variations caused by reflections from the surfaces
of the calibration loads and the windows of the calibration dewar.  
We then see from Equations \ref{multiplier}, \ref{Tsys}, and \ref{Taj2} 
that
\begin{equation}
T_{Aj}^{*} = {{D_{\mathrm{source}j} - D_{\mathrm{ref}j}}\over
{D_{\mathrm{ref}j} - D_{\mathrm{zero}j}}} \, \, T_\mathrm{sys}^{*} 
\label{Tastar}
\end{equation}
is the calibrated spectrum of the source.
This is the ``$(S-R)/R\,$'' method of data acquisition.  It provides
a high-quality spectrum of the source, because the gain of each
channel,
$ T_\mathrm{sys}^{*} /
(D_{\mathrm{ref}j} - D_{\mathrm{zero}j})$, 
is measured during the course of the observation.

In the AST/RO system, the data acquisition program  OBS
writes the values $T_{Aj}$ (usually calculated 
according to Equation \ref{Taj2}--- 
this is an option selectable by the observer)
as {\em scan} data, and writes $\eta_l$, $g_{\mathrm{s}}$, $El$,
$\langle T_\mathrm{sky}\rangle$,
$\langle T_\mathrm{rx}\rangle$, $T_\mathrm{surf}$ 
together with about fifty other
status variables, in the {\em scan header}.  
In the command {\tt gt} in the data reduction program COMB,
the multiplier, $M$, is calculated, and the data are converted to $T_{Aj}^*$.

The {\em forward spillover
and scattering efficiency}, $\eta_\mathrm{fss}$, is defined by
\citet{kutner81} to be the fraction of the telescope's forward
response which is also within the diffraction
pattern in and around the main beam.  
This quantity relates $T_A^{*}$ to the {\em radiation temperature scale}
$T_R^{*}$, which is the recommended temperature scale for
reporting millimeter-wave data \citep{kutner81}:
$T_R^{*} = T_A^{*}/\eta_\mathrm{fss} ~$.
The unusual optics of the AST/RO telescope (Figure \ref{fig:optics}) 
give it feed properties similar to that of prime-focus designs,
for which
$\eta_\mathrm{fss} \approx 1.0$ \citep{kutner81}.
When the
AST/RO beam, operating at 492 GHz, 
is pointed $4'$ from the Sun's limb, the excess
$T_A^{*}$ from the Sun is less than 1\% of the brightness of the
Sun's disk.
This indicates
$1.0 \geq \eta_\mathrm{fss} > 0.97$, and so for AST/RO,
$T_R^{*} = T_A^{*} \cdot (1.015 \pm 0.015)$.

\section{Conclusion}

AST/RO is the first submillimeter-wave telescope to operate 
in winter on the Antarctic Plateau.  During the first five years
of observing, the most serious operational difficulties have been
incapacitation of the single winterover scientist and lack of liquid
helium---these problems are being addressed through allocation of
additional resources and redundancy.  Site testing and observation
have demonstrated that South Pole is an excellent site having
good transparency and unusually low sky noise.  This observatory is
now a resource available to all astronomers on a proposal basis.

\acknowledgements

Rodney Marks died while serving as the year 2000 AST/RO winter-over
scientist, and it is to him that this paper is dedicated.
We thank Edgar Castro, Jeff Capara, Peter Cheimets, Jingquan Cheng, 
Robert Doherty, Urs Graf, James Howard, Gopal Narayanan, Maureen
Savage, Oliver Siebertz, and Volker Tolls
for their contributions to the project.  Jonas Zmuidzinas has generously
provided examples of his excellent SIS mixers.
We also thank Rick LeDuc and Bruce Bumble at JPL for making SIS 
junctions and Tom Phillips at Caltech 
for making them available to us.  
We thank Eric Silverberg and the Smithsonian Submillimeter Array 
Project for the optical guide telescope.
We thank Simon Radford of NRAO and Jeff Peterson of CMU for the data
shown in Figure \ref{fig:threesites}.
We thank Juan R. Pardo of Caltech for discussions on
atmospheric modeling and for carrying out
the calculations shown in Figure \ref{fig:pardo}.
The AST/RO group is grateful for the logistical support of the 
National Science Foundation (NSF), Antarctic Support Associates, 
Raytheon Polar Services Company,
and CARA during our polar expeditions.
The University of Cologne contribution to AST/RO was supported by
special funding from the Science Ministry of the Land
Nordrhein-Westfalen and by the Deutsche Forschungsgemeinschaft through
grant SFB 301.
This work was supported in part by  United States
National Science Foundation grant DPP88-18384, and by
the Center for Astrophysical Research in Antarctica and the NSF
under Cooperative Agreement OPP89-20223.

\bibliographystyle{apj}

\end{document}